\documentclass[showpacs,preprintnumbers,amssymb,nofootinbib,superscriptaddress,aps]{revtex4}
\usepackage{epsfig,color,wrapfig}
\usepackage{amsmath}
\usepackage{graphicx}
\usepackage{amssymb}
\usepackage{slashbox}
\usepackage{float}
\usepackage{bm}
\usepackage{dcolumn,multirow}
\usepackage[font={small}]{caption, subfig}
\usepackage{enumitem}
\setlist[itemize]{noitemsep,topsep=0pt}
\usepackage{mathtools}
\usepackage{pstricks}
\usepackage[toc,page]{appendix}
\usepackage{pifont}
\usepackage{braket}
\usepackage{bbold}
\usepackage{hyperref}

\usepackage{relsize}
\def\Babar{{\mbox{\slshape B\kern-0.1em{\smaller A}\kern-0.1em B\kern-0.1em{\smaller A\kern-0.2em R}}}}

\newcommand{\ba}{\begin{array}}
	\newcommand{\ea}{\end{array}}
\def\beq{\begin{equation}}
\def\eeq{\end{equation}}
\def\bea{\begin{eqnarray}}
\def\eea{\end{eqnarray}}
\def\nn{\nonumber}

\def\roughly#1{\mathrel{\raise.3ex\hbox
		{$#1$\kern-.75em\lower1ex\hbox{$\sim$}}}}

\def\gsim{\roughly>}
\def\sla#1{\raise.15ex\hbox{$/$}\kern-.57em #1}

\def\bd{B_d^0}

\def\order{\lower 1.8ex \hbox{\LARGE\~{}}}

\def\bd0tau{B\to D \tau\nu_{\tau}}

\def\be {\begin{equation}}
\def\ee {\end{equation}}

\usepackage[normalem]{ulem}

\definecolor{darkgreen}{cmyk}{1,0,1,0.4}

\definecolor{pink}{cmyk}{0.4,1,0.3,0}

\def\com2#1{\textcolor{red}{\it{#1}}}
\newcommand{\tcb}[1]{\textcolor{blue}{#1}}

\begin{document}
	
\title{New physics in $b\to s \ell\ell$ decays with complex Wilson coefficients }

\author{Aritra Biswas}
\affiliation{Indian Institute of Technology, North Guwahati, Guwahati 781039, Assam, India }

\author{Soumitra Nandi}
\affiliation{Indian Institute of Technology, North Guwahati, Guwahati 781039, Assam, India }

\author{Ipsita Ray}
\affiliation{Indian Institute of Technology, North Guwahati, Guwahati 781039, Assam, India }

\author{Sunando Kumar Patra}
\affiliation{Department of Physics, Bangabasi Evening College, 19 Rajkumar Chakraborty Sarani, Kolkata 700009, West Bengal, India }

\begin{abstract}  
	We perform a data-driven analysis of new physics (NP) effects in exclusive $b \to s \ell^+\ell^-$ decays in a model-independent effective theory approach with dimension six operators considering scalar, pseudo-scalar, vector and axial-vector operators with the corresponding Wilson coefficients (WC) taken to be complex. The analysis has been done with the most recent data while comparing the outcome with that from the relatively old data-set. We find that a left-handed quark current with vector muon coupling is the only one-operator $(\mathcal{O}_9)$ scenario that can explain the data in both the cases with real and complex WC with a large non-zero imaginary contribution. We simultaneously apply model selection tools like cross-validation and information-theoretic approach like Akaike Information Criterion (AIC) to find out the operator or sets of operators that can best explain the available data in this channel. The $\mathcal{O}_9$ with complex WC is the only one-operator scenario which survives the test. However, there are a few two and three-operator scenarios (with real or complex WCs) which survive the test, and the operator $\mathcal{O}_9$ is common among them.
\end{abstract}   

	\maketitle
	
\section{Introduction}
 In the last few years, the $b\to s\mu^+\mu^-$ decays have enjoyed a lot of attention both from the experimental and theoretical sides. These modes are potentially sensitive to new physics (NP) since the corresponding Standard Model (SM) contributions are loop suppressed. In $B(B_s) \to K^*(\phi) \mu\mu$ decays, plenty of NP-sensitive observables (viz. CP-averaged, CP-asymmetric and optimized angular observables) have been measured by different experimental collaborations like LHCb, Belle, ATLAS and CMS in different $q^2$-bins, where $q^2$ is the di-lepton invariant mass squared. A few angular observables have shown deviations from their respective SM predictions \cite{DescotesGenon:2012zf,Descotes-Genon:2013vna,Horgan:2013pva,Straub:2015ica}. The most interesting one among them is $P'_5$. LHCb, in 2015~\cite{Aaij:2015oid}, reported a tension at the level of 3.7$\sigma$ for $P'_5$ w.r.t the corresponding SM prediction. Very recently, LHCb have updated their results on CP-averaged angular observables with better statistics. The data on $P'_5$ still shows a deviation of $\sim$3$\sigma$ \cite{Aaij:2020nrf}. The source of these discrepancies could be the presence of one or more new interactions beyond the SM. On the other hand, it is also possible that the observed differences are due to poorly understood hadronic effects. Furthermore, these decay modes offer theoretically clean observables like $R_{K^{(*)}} = \frac{Br(B \to K^{(*)} \mu^+\mu^-)}{Br(B\to K^{(*)} e^+e^-)}$, which are useful to test lepton-flavor-universality violation (LFUV). The respective SM predictions are provided in refs. \cite{Hiller:2003js,Bordone:2016gaq}. The measurements on $R_K$ by LHCb and Belle are given in \cite{Aaij:2019wad} and \cite{Abdesselam:2019lab}, respectively. The most recent results on $R_{K^*}$ are available in refs. \cite{Aaij:2017vbb} (LHCb) and \cite{Abdesselam:2019wac} (Belle). These measurements are done in different $q^2$-bins. In our analysis, we will use the notation $R^{Low}_{K^*}$ and $R^{Central}_{K^*}$ from now on to represent $R_{K^*}$ corresponding to $q^2\in[0.045,1.1]{\rm GeV}^2$ and $[1.1,6]{\rm GeV}^2$, respectively. At the moment, the deviation between data and the corresponding SM predictions for these observables stand at the level of 2.5 to 3 $\sigma$. More precise measurements of these ratios might be unambiguous probes for NP. Different types of NP interactions (like vector, scalar etc.) may contribute to these decays and explain the data. Among the plethora of works present in the literature, here we will point out only a few model-independent studies based on the data available till Dec, 2019 \cite{Aebischer:2019mlg,Alok:2019ufo,Capdevila:2017bsm,Arbey:2018ics,Ciuchini:2019usw,Kowalska:2019ley,Bhattacharya:2019dot}. All these analyses have considered only real WCs.

In this article, we perform a model-independent analysis of NP affecting the $b\to s \ell^+\ell^-$ decay modes including all the data available till date with complex WC's. To the best of our knowledge, this is the first global model-independent analysis which does this. The operator basis is the same as that given in Ref.~\cite{Altmannshofer:2008dz}. Similar to ref. \cite{Bhattacharya:2019dot}, here too we find that ${\mathcal{O}}_{9} = \frac{e^2}{g^2} (\bar{s} \gamma_{\mu} P_L b)(\bar{\mu} \gamma^\mu \mu)$ is the only one operator scenario (for not only real but also complex WC) that can best explain the present data. It is thus tempting to look for other possible combinations of these operators with the potential to explain the data. However, a scenario with a large number of parameters can fit the observed data very well, but it suffers from the possibility of just fitting the noise and might hence lose sight of the important trends. The more important problem, therefore, is to optimize the number of parameters required to explain a certain observation \cite{Burnham, Geisser:1979}. To overcome this problem, we introduce penalized-likelihood information criteria, such as the sample-size-corrected Akaike Information Criterion (AIC$c$) \cite{akaike,CAVANAUGH1997201} which estimates the relative amount of information lost by a given model: the less the information lost by a model, the higher the quality of that model. For details, we refer to our earlier publications \cite{Bhattacharya:2016zcw,Bhattacharya:2018kig,Bhattacharya:2019dot} and the references therein. Also, the most generally applicable, powerful, and reliable method for comparison of the predictive capability of a model, (although computationally expensive) is `cross-validation'\cite{Andrae:2010gh}. In accordance to our earlier publication \cite{Bhattacharya:2019dot}, we use both AIC$c$ and cross-validation to pin down the best possible scenarios.

\section{Theory}

 At the low-energy scale ($\mu \approx m_b$), the effective Hamiltonian and the operator basis for exclusive $b\to s \mu^+\mu^-$ decays has been taken from \cite{Bobeth:1999mk, Altmannshofer:2008dz, Altmannshofer:2014rta} and is written as:

	\begin{equation}
		{\cal H}_{eff} =  \frac{4\,G_F}{\sqrt{2}} \sum_{q=u,c}
		 \lambda_q \Big( C_1 \mathcal O^q_1 + C_2 \mathcal O^q_2  + \sum_{i=3,..,6,P,S} (C_i \mathcal O_i + C'_i \mathcal O'_i) + \sum_{i=7,..,10} (\tilde{C_i} \tilde{\mathcal O}_i + {\tilde C}^{\prime}_i \tilde{\mathcal O}'_i)\Big)\,,
		\label{eq:Heff}
	\end{equation}

with the CKM combination $\lambda_q=V_{qb}V_{qs}^*$, and the WC corresponding to the operators $\mathcal O_i$ and $\mathcal O'_i$ are given by $C_i$ and $C'_i$, respectively. The set of tilde-operators are as given below:

	\begin{align}
		\nn {\tilde{\mathcal{O}}}_{7} &= \frac{e}{g^2} m_b (\bar{s} \sigma_{\mu \nu} P_R b) F^{\mu \nu}, ~~~~~~~~~~~~~~~~~
		{\tilde{\mathcal{O}}_{7}^\prime} = \frac{e}{g^2} m_b (\bar{s} \sigma_{\mu \nu} P_L b) F^{\mu \nu},\\ 
		\nn {\tilde{\mathcal{O}}}_{8} &= \frac{1}{g} m_b (\bar{s} \sigma_{\mu \nu} T^a P_R b) G^{\mu \nu a}, ~~~~~~~~~~~~~~
		{\tilde{\mathcal{O}}_{8}^\prime} = \frac{1}{g} m_b (\bar{s} \sigma_{\mu \nu}T^a P_L b) G^{\mu \nu a},\\
		\nn {\tilde{\mathcal{O}}_{9}} &= \frac{e^2}{g^2} (\bar{s} \gamma_{\mu} P_L b)(\bar{\mu} \gamma^\mu \mu), ~~~~~~~~~~~~~~~~~~~ {\tilde{\mathcal{O}}_{9}^\prime} = \frac{e^2}{g^2} (\bar{s} \gamma_{\mu} P_R b)(\bar{\mu} \gamma^\mu \mu), \\
		 {\tilde{\mathcal{O}}_{10}} &=\frac{e^2}{g^2} (\bar{s}  \gamma_{\mu} P_L b)(  \bar{\mu} \gamma^\mu \gamma_5 \mu), ~~~~~~~~~~~~~~~ {\tilde{\mathcal{O}}_{10}^\prime} =\frac{e^2}{g^2} (\bar{s}  \gamma_{\mu} P_R b)(  \bar{\mu} \gamma^\mu \gamma_5 \mu).
		\label{eq:effoprsm}
	\end{align}

In the above equation, $g$ is the strong coupling constants. The current-current and the QCD-penguin operators are taken from \cite{Bobeth:1999mk}. Though the primed operators and $\mathcal{O}_{S,P}$ vanish or are highly suppressed in the SM, they could be relevant in some NP scenarios. For a proper organization of the perturbative expansion of the WCs, a normalization factor $1/g^2$ has been introduced in front of the operators $\tilde{\mathcal{O}}_i$ ($i=7,..,10$) \cite{Bobeth:1999mk}. To be consistent, a similar factor has been introduced in front of the primed operator. Note that we are treating the new WCs as free parameters and fitting them from data. Therefore, the discussion on this choices of a proper normalization is more relevant in the context of the SM operators and the corresponding WCs, in particular, in their renormalization group equation (RGE) evolution. For these operators and the corresponding WCs, it is possible to change the normalization to another commonly used basis \cite{Altmannshofer:2014rta,Descotes-Genon:2015uva}: 
\begin{equation}
\mathcal{O}_i^{(\prime)} = \frac{g^2}{16 \pi^2}  \tilde{\mathcal{O}_i}^{(\prime)}\ \ \ \text{and} \ \ \  {C}_i^{(\prime)} = \frac{16 \pi^2}{g^2}  \tilde{{C}_i}^{(\prime)}.
\label{eq:basis}
\end{equation}
We can see that both the basis of operators are equivalent. With a change in the normalization of the operators, there will be an appropriate scaling in the respective WCs. As mentioned above, since we fit the new WCs from data, working in any basis should be fine with the proper scaling of the corresponding WCs. In our analysis, we have considered the NP effects in the following set of operators (non-tilde basis):
 	\begin{align}
 	{\mathcal{O}}_{7}^\prime &= \frac{e}{16 \pi^2} m_b (\bar{s} \sigma_{\mu \nu} P_L b) F^{\mu \nu},~~~~~~~~~~~~~~~~
 	\nn \mathcal{O}_{9} = \frac{e^2}{16 \pi^2} (\bar{s} \gamma_{\mu} P_L b)(\bar{\mu} \gamma^\mu \mu), ~~~~~~~~~~~~~~ \mathcal{O}_{9}^\prime = \frac{e^2}{16\pi^2} (\bar{s} \gamma_{\mu} P_R b)(\bar{\mu} \gamma^\mu \mu), \\
 	\nn \mathcal{O}_{10} &=\frac{e^2}{16 \pi^2} (\bar{s}  \gamma_{\mu} P_L b)(  \bar{\mu} \gamma^\mu \gamma_5 \mu), ~~~~~~~~~~~~~ \mathcal{O}_{10}^\prime =\frac{e^2}{16 \pi^2} (\bar{s}  \gamma_{\mu} P_R b)(  \bar{\mu} \gamma^\mu \gamma_5 \mu),~~~~~~~~~
 		\nn {\mathcal{O}}_{S}  =\frac{e^2}{16\pi^2} m_b (\bar{s} P_R b)(  \bar{\mu} \mu),\\
 		{\mathcal{O}}_{S}^\prime &=\frac{e^2}{16\pi^2} m_b (\bar{s} P_L b)(  \bar{\mu} \mu) ,~~~~~~~~~~~~~
 		\qquad {\mathcal{O}}_{P} =\frac{e^2}{16\pi^2} m_b (\bar{s} P_R b)(  \bar{\mu} \gamma_5 \mu) , ~~~~~~~~~~~~ {\mathcal{O}}_{P}^\prime =\frac{e^2}{16\pi^2} m_b (\bar{s} P_L b)(  \bar{\mu} \gamma_5 \mu)\,.
 	\label{eq:effopr}
 	\end{align}
  The relevant WCs are the following: $C_7^{\prime}$, $\Delta C_9$, $C_9^{\prime}$, $\Delta C_{10}$, $C_{10}^{\prime}$, $C_S^{(\prime)}$, and $C_P^{(\prime)}$. 
  
The amplitude and the corresponding rate distributions in $B\to K^{(*)}\ell\ell$ decays are defined using the hadronic form factors computed in full QCD. Our treatment of the decay amplitude in $B\to K\ell\ell$ is closely related to that given in ref. \cite{Altmannshofer:2014rta} and that for the $B\to K^*\ell\ell$ decays are taken from \cite{Altmannshofer:2008dz}. The SM WCs ${\tilde C}_7$, ${\tilde C}_9$ and ${\tilde C}_{10}$ are taken in next-to-next-to-leading logarithmic (NNLL) approximation, the details of which are provided in \cite{Bobeth:1999mk}. The contributions from the four-quark current-current operators $\mathcal O^q_1$, $\mathcal O^q_2$ (with $q=u$ and $c$) and quark-penguin operators $\mathcal O_{3..6}$ are included in an effective coefficient ${\tilde C}_{9}^{ eff}$ of the operator $\mathcal O_9$ which also includes the leading order (LO) charm-loop effects. Note that ${\tilde C}_{9}^{ eff}$ is in general complex. A part of the contributions from the operators $\mathcal O_{3..6}$ are also included in ${\tilde C}_{7}^{eff}$. The perturbatively calculable non-factorizable corrections to the operators $\mathcal O_1$ and $\mathcal O_2$ at order $\alpha_s$ are taken from \cite{Asatrian:2001de,Asatryan:2001zw} and added as a correction to ${\tilde C}_{9}^{ eff}$. Other non-factorizable corrections which have been considered in this analysis include the weak annihilation and the hard spectator scattering contributions as given in \cite{Beneke:2000wa,Beneke:2001at}. Note that both these additional contributions include the charm loop effects, and they are complex. In this analysis, we have not considered the contributions from the non-factorizable power corrections which are expected to be very small as pointed out in a very recent analysis \cite{Gubernari:2020eft}. When the final state contains a vector meson, one can construct various helicity/transversity amplitudes like $A_{\bot L,R}$, $A_{\| L,R}$, $A_{0L,R}$ etc., which are then used to form angular coefficients relevant in defining the CP-symmetric and asymmetric observables measured by the different experimental collaborations. As mentioned earlier, details about various transversity amplitudes and the respective angular coefficients can be obtained from \cite{Altmannshofer:2008dz}. Here, the non-factorizable corrections are considered as additional contributions $\Delta A_{\bot L,R}$, $\Delta A_{\| L,R}$ and $\Delta A_{0 L,R}$ to the transversity amplitudes for details see section 4.3 of ref. \cite{Altmannshofer:2008dz}. In the SM, the complex phases will appear in all the $A_{i L,R}$'s and $\Delta A_{i L,R}$'s. 

The NP contributions to operators $\mathcal{O}_{9,10}$ are given by $\Delta C_{9,10}$.  We have not explored the possibility of new physics effects in $\mathcal{O}_{7}$ which is tightly constrained from the available data on inclusive and exclusive radiative decays. However, we have considered new physics effects in the chirality flipped operator $\mathcal{O'}_{7}$, the real part of the corresponding WC $C'_7$ is also tightly constrained by the data. However, as pointed out in \cite{Altmannshofer:2012az}, the direct CP-asymmetry $A_{CP}$ in the inclusive $b\to s\gamma$ decays tightly constrained the imaginary part of $ C_7$ along with it's real part, though it has marginal effect in the $Im(C'_7)$. Note that in our analysis, we are not using $A_{CP}(b\to s\gamma)$ as an input since both its predicted and the measured values have large error. We have constrained both the real and imaginary parts of the $C'_7$ from all  other available data as will be discussed below, and we do not expect the currently available inputs on $A_{CP}(b\to s\gamma)$ to provide tighter constraints on $Im(C'_7)$.

\section{Experimental Inputs}

\begin{table}[t]
	\begin{center}
		\begin{tabular}{|c|c|c|c|c|}
			\hline
			Observables  &  LHCb 2016 \cite{Aaij:2015oid} &  LHCb 2020 \cite{Aaij:2020nrf} & pull wrt LHCb 2016 &pull wrt LHCb 2020\\
			\hline
			$A_{FB}^{[0.1,0.98]}$ & $-$0.003 $\pm$ 0.058 & $-$0.004 $\pm$ 0.04 & 1.43 & 2.03\\
			
			$S_3^{[1.1,2.5]}$ & $-$0.077 $\pm$ 0.096 & $-$0.107 $\pm$ 0.052 & $-$0.82 & $-$2.08\\
			
			$S_7^{[1.1,2.5]}$ & $-$0.219 $\pm$ 0.099 & $-$0.107 $\pm$ 0.063 & $-$2.31 & $-$1.86\\
			
			$S_8^{[1.1,2.5]}$ & $-$0.098 $\pm$ 0.116 & $-$0.174 $\pm$ 0.075 & $-$0.9 & $-$2.4\\
			
			$S_9^{[1.1,2.5]}$ & $-$0.119 $\pm$ 0.096 & $-$0.112 $\pm$ 0.054 & $-$1.25 & $-$2.07\\
			
			$P_1^{[1.1,2.5]}$ & $-$0.451 $\pm$ 0.579 & $-$0.617 $\pm$ 0.297 & $-$0.8 & $-$2.12\\
			
			$P_3^{[1.1,2.5]}$ & 0.350 $\pm$ 0.292 & 0.324 $\pm$ 0.148 & 1.2 & 2.2\\
			
			$P_6^{'\hspace{0.05 cm}[1.1,2.5]}$ & $-$0.463 $\pm$ 0.212 & $-$0.226 $\pm$ 0.128 & $-$2.29 & $-$1.94\\
			
			$P_8^{'\hspace{0.05 cm}[1.1,2.5]}$ & $-$0.208 $\pm$ 0.248 & $-$0.366 $\pm$ 0.158 & $-$0.89 & $-$2.4\\
			
			$S_5^{[4, 6]}$ & $-$0.146 $\pm$ 0.078 & $-$0.204 $\pm$ 0.053 & 2.41 & 2.43\\
			
			$S_7^{[4, 6]}$ & $-$0.016 $\pm$ 0.081 & $-$0.136 $\pm$ 0.053 & $-$0.26 & $-$2.66\\
			
			$P_2^{[4, 6]}$ & 0.042 $\pm$ 0.088 & 0.105 $\pm$ 0.069 & $-$2.65 & $-$2.48 \\
			
			$P_5^{'\hspace{0.05 cm}[4, 6]}$ & $-$0.300 $\pm$ 0.160 & $-$0.439 $\pm$ 0.117 & 2.96 & 2.85\\
			
			$P_6^{'\hspace{0.05 cm}[4, 6]}$ & $-$0.032 $\pm$ 0.167 & $-$0.293 $\pm$ 0.117 & $-$0.26 & $-$2.6\\
			
			$S_7^{[1.1, 6]}$ & $-$0.077 $\pm$ 0.050 & $-$0.09 $\pm$ 0.034 & $-$1.69 & $-$2.86\\
			
			$P_6^{'\hspace{0.05 cm}[1.1, 6]}$ & $-$0.166 $\pm$ 0.11 & $-$0.197 $\pm$ 0.076 & $-$1.66 & $-$2.83\\
			
			
			\hline
		\end{tabular}
		\caption{List of a few observables with pulls $> 2$ of the `Likelihood dataset 2020' and `Likelihood dataset 2016'. The superscripts on the observables indicate the $q^2$ range in {\it GeV}$^2$. The corresponding SM predictions can be seen from \cite{Altmannshofer:2014rta,Bhattacharya:2019dot}.  }
		\label{tab:pullsliklihood}
	\end{center}
\end{table}

\begin{table}[htbp]
	\begin{center}
	\begin{tabular}{|c|c|c|c|c|}
		\hline
			\multicolumn{5}{|c|}{List of few more observables with pull $>$ 2 which are common to all the datasets} \\
		\hline
		\multicolumn{2}{|c|}{Observables} & \multicolumn{2}{|c|}{Measured values}  & \multicolumn{1}{c|}{Respective pulls} \\
		\hline 
		\multicolumn{2}{|c|}{$BR(B^{0}\to K^{0}\mu^+\mu^-)^{[1, 6]}$ Belle \cite{Abdesselam:2019lab}} & \multicolumn{2}{|c|}{$(3.1\pm 1.9)\times 10^{-8}$}   & \multicolumn{1}{c|}{ $-$3.9}\\
		\multicolumn{2}{|c|}{	$A_I(B\to K \mu^+\mu^-)^{[1,6]}$  Belle \cite{Abdesselam:2019lab}} & \multicolumn{2}{|c|}{$-0.52\pm 0.19$} & \multicolumn{1}{c|}{ $-$2.8}\\
		\multicolumn{2}{|c|}{${P_4^{'}(B^0\to K^{*0}\mu^+\mu^-)}^{[4,6]}$  ATLAS \cite{Aaboud:2018krd}} & \multicolumn{2}{|c|}{$0.64\pm 0.38$}  &\multicolumn{1}{c|}{  2.97}\\
		\multicolumn{2}{|c|}{$S_4(B^0\to K^{*0}\mu^+\mu^-)^{[4, 6]}$  ATLAS \cite{Aaboud:2018krd} } & \multicolumn{2}{|c|}{$0.32\pm 0.18$ } &\multicolumn{1}{c|}{ 2.87}\\	
		\multicolumn{2}{|c|}{$F_H(B^+\to K^+\mu^+\mu^-)^{[2,4.3]}$  CMS \cite{Sirunyan:2018jll}} & \multicolumn{2}{|c|}{$0.85\pm 0.35$}  & \multicolumn{1}{c|}{  2.34}\\
		\hline
	\end{tabular}
	\caption{The data with pulls $> 2$ which are common to all the datasets. The superscripts on the observables indicate the $q^2$ range in {\it GeV}$^2$.}
	\label{tab:pullcommon}
\end{center}
\end{table}

In what follows, we categorically present the experimental inputs in our analysis. Our main analysis is based on the following data sets: 
 
\begin{itemize}
	\item \textbf{Likelihood dataset 2020:}
 \begin{enumerate}[label=(\roman*)]
   \item  Measured values of the angular observables in $B^0\to K^{*0}\mu^+\mu^-$ decays in different bins, gathered from ref.~\cite{Aaij:2020nrf} (LHCb).
   
	\item For $B^{(0,+)}\to K^{(0,+)}\mu^+\mu^-$ decays, we consider: (i) The LHCb and Belle measurements on the  (differential) branching fractions in different bins which are taken from \cite{Aaij:2014pli} and \cite{Abdesselam:2019lab}, respectively. (ii) Isospin asymmetries measured by Belle \cite{Abdesselam:2019lab} and LHCb~\cite{Aaij:2014pli}. (iii) Binned data on the angular observables for $B^+\to K^+\mu^+\mu^-$ ($A_{FB}$ and $F_H$) obtained from CMS (ref.~\cite{Sirunyan:2018jll}). (iv) The inputs on $R_K$ have been taken from refs.~\cite{Aaij:2019wad} (LHCb) and \cite{Abdesselam:2019lab} (Belle) where the Belle results for different bins are also included. We do not use the isospin averaged measuement for $R_K$ by Belle and include the separate estimates given for $R_{K^{(0,+)}}$ instead.
	
	\item For $B\to K^*\mu^+\mu^-$ decays, we include: (i) Binned data on the differential branching fraction from LHCb for $B^0\to K^{*0}\mu^+\mu^-$ and $B^+\to K^{*+}\mu^+\mu^-$ decays, taken from ref.~\cite{Aaij:2016flj} and \cite{Aaij:2014pli}, respectively. (ii) The measured values of the angular observables in $B^0\to K^{*0}\mu^+\mu^-$ decays in different bins gathered from \cite{Aaboud:2018krd} (ATLAS). (iii) The measured values of $P_4^\prime$ and $P_5^\prime$ for $B^0\to K^{*0}\mu^+\mu^-$ by Belle in ref.~\cite{Wehle:2016yoi}. (iv) Isospin-asymmetry measurements corresponding to $B\to K^*\mu^+\mu^-$ from LHCb (ref.~\cite{Aaij:2014pli}). Furthermore, the measured values of $R_{K^{*(+,0)}}$ in different bins are taken from refs.~\cite{Aaij:2017vbb} and \cite{Abdesselam:2019wac}, respectively. 
	
	\item Binned data on the differential branching fractions and angular observables (CP-averaged and asymmetric) for $B_s\to\phi\mu^+\mu^-$ (LHCb) from ref.~\cite{Aaij:2015esa}. Other inputs are: (i) $\text{BR}(B\to X_s\gamma)_{E_\gamma>1.6\,\text{GeV}}$~\cite{Misiak:2017bgg}, (ii) $\text{BR}(B^{+/0}\to K^*\gamma)$~\cite{Amhis:2014hma},(iii) 	$\overline{\text{BR}}(B_s\to \phi\gamma)$~\cite{Aaij:2012ita}, (iv) $Br(B_s \to \mu\mu)$~\cite{hflav}.
\end{enumerate} 
\end{itemize} 
For the purpose of comparison we perform our analysis over two other datasets apart from the set mentioned above, which we call the (i) Moment 2016 dataset, (ii) Likelihood 2016 dataset, which are as given below
\begin{itemize}
	\item \textbf{Moments dataset 2016:} Values for the angular observables measured by the ``Method of moments" in $B^0\to K^{*0}\mu^+\mu^-$ decays in different bins, gathered from refs.~\cite{Aaij:2015oid} (LHCb).
	\item \textbf{Likelihood dataset 2016:} Values for the angular observables due to the ``Unbinned Maximum likelihood" in $B^0\to K^{*0}\mu^+\mu^-$ decays in different bins, gathered from refs.~\cite{Aaij:2015oid} (LHCb).
\end{itemize}
The only difference between the datasets mentioned above is subject to the angular observables in the $B^0\to K^{*0}\mu^+\mu^-$ sector due to LHCb. The inputs discussed in items (ii), (iii) and (iv) of our primary dataset (\textbf{Likelihood dataset 2016}) are included in the \textbf{Moments dataset 2016} and \textbf{Likelihood dataset 2016} as well. Beyond the fit scenarios discussed above, we have also carried out an analysis without considering the inputs on CP-asymmetric observables in $B_s\to \phi\mu\mu$ decays.  
 
In addition to other inputs (e.g. CKM matrix elements) \cite{CKMfitter}, we have included the lattice input $f_{B_s} = 0.2284 \pm 0.0037$ {\rm~ GeV}~\cite{Aoki:2019cca} in all the fits. Unless otherwise specified, all numerical uncertainties quoted in this analysis denote the $1 \sigma$ (68.27\% c.l.) range. Note that in \cite{Aaij:2020nrf}, LHCb did not update the measurement on CP-asymmetric observables in $B^0\to K^{*0}\mu^+\mu^-$ decays.  Therefore, to check whether they provide tighter constraints on the complex WCs, we incorporate those measurements \cite{Aaij:2015oid} in a different fit along with the data defined as \textbf{Likelihood dataset 2020}. This additional set of $B^0\to K^{*0}$ asymmetric observables is completely uncorrelated with the Likelihood dataset 2020. Also, we consider the bins for $q^2\leq 6$ GeV$^2$ to avoid any contamination from the charm resonances. Other relevant information about the inputs from theory can be obtained from our earlier publication \cite{Bhattacharya:2019dot} and the references therein.  

\section{Analysis and Results} 

Our `Likelihood dataset 2020' contains a total of 224 observables. We first check whether each of the new operators defined in eq.\ref{eq:effopr} can independently explain the present data after including all these observables. To do so, we perform a frequentist statistical analysis optimizing a $\chi^2$ statistic which is a function of the relevant WCs. Separate covariance matrices are constructed for statistical (systematic) uncertainties where necessary. In the post-process for each fit, we obtain the corresponding fit-quality using $p$-value. For other technical details about the fit, we refer to our previous study \cite{Bhattacharya:2019dot}. We perform separate analyses with complex, as well as real WCs.

For the analysis with all data, we obtain a very poor quality fit for all the one operator scenarios with the respective $p$-values $<<$ 1\%. It is hard to get a statistically significant result for the NP to fit all the 224 data points that we consider. There must be some data points that can not be explained alongside the rest of the data. 
Note that $\mathcal{O}_9$ is the only one operator scenario in both the analyses with complex and real WCs having a relatively better $p$-value, though it is $< 1$\%. Table \ref{tab:oneopralldata} shows the results of the fits corresponding to $O_9$ only. In all other one operator scenarios, the fit quality is very poor, with the respective $p$-values $\approx 0$. Parameter uncertainties are obtained from both the Fisher-matrix and the 1-dimensional (1-D) profile-likelihood of the parameter of interest. We have repeated the fit with the new `Likelihood 2020 datasets' in two more subsets: (i) without considering the inputs from the CP-asymmetric observables in $B_s \to\phi\mu\mu$, (ii) including the inputs from the CP-asymmetric observables in $B\to K^*\mu\mu$ from `Likelihood 2016 datasets'. For comparison, the results for these different datasets corresponding to the one-operator scenario $O_9$ are presented in the same table \ref{tab:oneopralldata}. The allowed confidence intervals for $Re(\Delta C_9)$ are consistent with each other for all the three different fits, and large imaginary contributions are allowed by the present data even after dropping the CP-asymmetric observables in $B_s \to\phi\mu\mu$ and $B\to K^*\mu\mu$ from the fit. We have given the values of the respective $\chi^2_{SM}$ and mentioned the corresponding $p$-values (in SM) for all the three different scenarios. For all these cases, following ref. \cite{Capdevila:2018pull} we have calculated Pull$_{SM}$ which is defined as     
\begin{equation}
\text{Pull}_{\text{SM}} = \sqrt{2} \text{Erf}^{-1} [F(\Delta{\chi^2_{\text{SM}};n_{\text{dof}}})], \ \ \ \text{with}\  \Delta{\chi^2_{\text{SM}}} = \chi^2_{SM}-({\chi^2_{min}})_{NP}.
\label{eq:pullsmnp}
\end{equation}
Here, $F$ is the $\chi^2$ cumulative distribution function and $n_{\text{dof}}$ is the associated number of degrees of freedom. In this case we are comparing the respective NP scenarios against the SM by defining Pull$_{SM}$. A large value of this quantity indicates a large deviation from the SM. The respective values of the Pull$_{SM}$ are given in the third column of the above table.

\begin{table}[t]
	\small
	\begin{ruledtabular}
		\renewcommand*{\arraystretch}{1.2}
		\begin{tabular}{cccc}
			
			\multicolumn{4}{c|}{Fit scenario: Likelihood dataset 2020 ($\chi^2_{SM}= 392.7$,  $p$-value $(\%)$ in SM = $2.3\times 10^{-9}$)}    \\
			\hline
			$\chi _{Min}^2 / DOF $  &  $p$-value $(\%)$ &  Pull$_{SM}$ & confidence intervals \\
			\hline
			302.4/223 & 0.03 & 9.5 & $Re(\Delta C_9)\to -1.11 \pm 0.11 $  \\
			297.5/222 & 0.05 &  9.5 & $Re(\Delta C_9)\to -1.16 \pm 0.11$, $Im(\Delta C_9)\to 1.38 \pm 0.34$  \\
			\hline
			\multicolumn{4}{c}{Likelihood dataset 2020: without CP-asymmetric observables in $B_s \to \phi\mu\mu$ ($\chi^2_{SM}= 384.4$,  $p$-value $(\%)$ in SM = $4.2\times 10^{-10}$)  } \\
			\hline
			 $\chi _{Min}^2 / DOF $ &  $p$-value $(\%)$  &  Pull$_{SM}$ & confidence intervals \\
			 294.1/211 & 0.01 & 9.5 & $Re(\Delta C_9)\to -1.11 \pm 0.11 $ \\
			 289.4/210 & 0.02 & 9.5 & $Re(\Delta C_9)\to -1.17 \pm 0.12$, $Im(\Delta C_9)\to -1.47 \pm 0.34$ \\ 
			 \hline
			\multicolumn{4}{c}{Fit scenario: Likelihood dataset 2020 $+$ CP-asymmetric observables in $B \to K^*\mu\mu$ from LHCb Likelihood dataset 2016}\\
				\multicolumn{4}{c}{ ($\chi^2_{SM}= 425.9$,  $p$-value $(\%)$ in SM = $2.8\times 10^{-8}$)}\\
			$\chi _{Min}^2 / DOF $ &  $p$-value $(\%)$  &  Pull$_{SM}$ & confidence intervals \\
			\hline
			335.6/258 &  0.08 & 9.5 & $Re(\Delta C_9)\to -1.11 \pm 0.11$ \\
		333.0/257 & 0.10 &9.4 & $Re(\Delta C_9)\to -1.15 \pm 0.11$,~~~~$Im(\Delta C_9) \to -1.26 \pm 0.40$ 
		\end{tabular}
	\end{ruledtabular}
	\caption{\small Fit results for the frequentist analysis with all data in one-operator($\mathcal{O}_9$) scenarios with real and complex WCs. The cases without the CP-asymmetric observables in $B_s \to \phi\mu\mu$, and with the CP-asymmetric observables in $B \to K^*\mu\mu$ are presented separately.}
	\label{tab:oneopralldata}
\end{table}

In the following we will discuss the possibility of improvement in the statistical significance of the fits. To proceed further, we have computed the pulls of the SM estimates w.r.t the corresponding measured values for all the observables that make up the different datasets as discussed above. The pull corresponding to the $i^{th}$ observable $\mathcal{O}_i$ is defined as:
\begin{equation}
pull_i = \frac{\mathcal{O}_i^{exp} - \mathcal{O}_i^{SM}}{\Delta \mathcal{O}_i}
\label{eq:pullsm}
\end{equation}
where $\Delta \mathcal{O}_i$ corresponds to the uncertainty of the data, including theoretical uncertainties. The SM predictions are provided in \cite{Altmannshofer:2014rta}. From our analysis, we found a few angular observables (listed in table \ref{tab:pullsliklihood}) whose measured values (from LHCb) have pulls greater than 2 w.r.t the corresponding SM estimates for the likelihood dataset 2020. The table also shows the pulls corresponding to the likelihood dataset 2016 for the same observables so that the trend of the data and the differences between the two sets become clear. The current measured values of a few observables have larger pulls than  the corresponding measured values from the 2016 dataset. The measured observables with pulls $>$ 2 from other experimental collaborations like ATLAS, CMS ad Belle have been shown in table \ref{tab:pullcommon}. A similar analysis has been carried out for the LHCb moments dataset (2016) which shows that there are only a few observables for which the measured values have a pull $>$ 2. They are $A_4^{[5,6]}$, $S_3^{[4,5]}$ and $S_4^{[2,3]}$, respectively.  

Besides this, using results of fit to all data, we compare the fitted values (in an NP scenario ) of each observable to their measured values by defining the pull, which for the $i^{th}$ observable is given by 
\begin{equation}
pull_i^{NP} = \frac{\mathcal{O}_i^{exp} - \mathcal{O}_i(C_k^{NP})}{\Delta \mathcal{O}_i}.
\label{eq:pullNP}
\end{equation} 
Here, $\mathcal{O}_i(C_k^{NP})$ is the predicted value of the $i^{th}$ observable in a new physics scenario with the best fit value of $C_k^{NP}$. A few observations from the pull analyses of eq. \ref{eq:pullNP}:
\begin{enumerate}\label{item:disclist1}
	\item Apart from $\mathcal{O}_9$, in all the other one-operator NP scenarios, the data points listed in tables \ref{tab:pullsliklihood} and \ref{tab:pullcommon} with a pull $>$ 2 are also selected with a pull$^{NP} > $ 2.
	
	\item Only for the one-operator scenario $\mathcal{O}_9$, the three data points: $P_5^{'\hspace{0.05 cm}[4, 6]}(LHCb)$, $S_5^{[4, 6]}(LHCb)$, and $P_2^{[4, 6]}(LHCb)$ have pull$^{NP} < $ 2. 
	
	\item Note that apart from the data points given in table \ref{tab:pullsliklihood} and \ref{tab:pullcommon}, the LFUV data points $R_{K^{*0}}^{[0.045,1.1]}(LHCb)$, $R_{K^{*0}}^{[1.1,6]}(LHCb)$, and $R_{K}^{[1.1,6]}(LHCb)$ also have pull $> 2$ compared to their SM estimates. However, in different NP scenarios, the estimates of pull$^{NP}$ for $R_{K^{*0}}^{[1.1,6]}(LHCb)$ and $R_{K}^{[1.1,6]}(LHCb)$ are $< 2$, though, pull$^{NP} > 2$ for $R_{K^{*0}}^{[0.045,1.1]}(LHCb)$ in various NP scenarios including $\mathcal{O}_9$. As will be discussed later, it is hard to explain the data on $R_{K^{*0}}^{[0.045,1.1]}(LHCb)$ in the NP scenarios that we are considering in our analysis. We have included all these important data points on LFUV observables in our fits.
	
	\item  In all the NP scenarios, a few additional data points are selected with a pull$^{NP} \gsim $2. The data points which are common among all of them are the following: $P_5^{'}(B^0\to K^{*0}\mu^+\mu^-)^{[4, 6]}$ (ATLAS), $S_5(B^0\to K^{*0}\mu^+\mu^-)^{[4, 6]}$ (ATLAS), $\frac{dB}{dq^2}(B^{+}\to K^{+}\mu^+\mu^-)^{[0.1, 0.98]}$ (LHCb) and $BR(B^{+}\to K^{+}\mu^+\mu^-)^{[1,6]}$ (Belle). 
	
	\item Note that in the one-operator scenario $\mathcal{O}_9$, the four data points mentioned in item 4 are the only additional points which are picked up with a pull$^{NP} > $ 2 in addition to what we have discussed earlier in points 1 and 2.
	
	\item For the one operator scenarios other than $\mathcal{O}_9$ the number of data points with pull$^{NP} > $ 2 are more than the four data points mentioned in point 3.
	
	\item \label{fn1}Note that the data points: $P_5^{'}(B^0\to K^{*0}\mu^+\mu^-)^{[4, 6]}$ (ATLAS), $S_5(B^0\to K^{*0}\mu^+\mu^-)^{[4, 6]}$ (ATLAS), $\frac{dB}{dq^2}(B^{+}\to K^{+}\mu^+\mu^-)^{[0.1, 0.98]}$ (LHCb), and $BR(B^{+}\to K^{+}\mu^+\mu^-)^{[1,6]}$ (Belle)  (which are common to all other datasets) are picked up with a pull$^{NP} >$ 2 in the analyses with other datasets.
\end{enumerate}

\begin{table}[t]
	\begin{center}
		\begin{tabular}{c|c}
			\hline
				\multicolumn{2}{c}{Observables in \textbf{List-1} with the respective pulls in the one-operator scenario $\mathcal{O}_9$}  \\
			\hline 
			$BR(B^{0}\to K^{0}\mu^+\mu^-)^{[1, 6]}$ (Belle) $\rightarrow$ -2.92 & $S_7^{[4, 6]}$ (LHCb) $\rightarrow$ - 2.56  \\
			${P_4^{'}(B^0\to K^{*0}\mu^+\mu^-)}^{[4,6]}$  (ATLAS) $\rightarrow$ 2.90 & $P_6^{'\hspace{0.05 cm}[4, 6]}$ (LHCb) $\rightarrow$ - 2.50 \\
			$S_4(B^0\to K^{*0}\mu^+\mu^-)^{[4, 6]}$  (ATLAS) $\rightarrow$ 2.86 & $BR(B^{+}\to K^{+}\mu^+\mu^-)^{[1,6]}$ (Belle)$\rightarrow$ 2.39  \\
			$A_I(B\to K \mu^+\mu^-)^{[1,6]}$  (Belle) $\rightarrow$ -2.81 & $\frac{dB}{dq^2}(B^{+}\to K^{+}\mu^+\mu^-)^{[0.1, 0.98]}$ (LHCb)$\rightarrow$   2.18 	 \\
			$S_7^{[1.1, 6]}$  (LHCb) $\rightarrow$ - 2.64 & $S_5(B^0\to K^{*0}\mu^+\mu^-)^{[4, 6]}$ (ATLAS)$\rightarrow$  2.10    \\
				$P_6^{'\hspace{0.05 cm}[1.1, 6]}$ (LHCb) $\rightarrow$ - 2.61 & $P_5^{'}(B^0\to K^{*0}\mu^+\mu^-)^{[4, 6]}$ (ATLAS) $\rightarrow$   2.02 \\	
			\hline
		\end{tabular}
		\caption{The respective values of pull$^{NP}$ for the observables in \textbf{List-1} which are obtained using eq. \ref{eq:pullNP} for the one-operator scenario $\mathcal{O}_9$ with complex WC. For the scenario with real WC also, these observables have pull$^{NP}$ $> 2$. The superscripts on the observables indicate the $q^2$ range in {\it GeV}$^2$.}
		\label{tab:pullNP}
	\end{center}
\end{table}

Following the above  discussion and with the primary intention to get a fit with allowed $p$-values, we prepare a list of data as given below which we will drop while fitting,
\begin{itemize}
	\item \textbf{List-1}: This list is prepared from the \textbf{Likelihood dataset 2020}. It contains the observables listed in tables \ref{tab:pullsliklihood} and \ref{tab:pullcommon} for the same dataset with a pull $>$ 2.5. As discussed above, the optimised observable $P_5^{'\hspace{0.05 cm}[4, 6]}$ has not been included in this list. Also, it includes $P_5^{'}(B^0\to K^{*0}\mu^+\mu^-)^{[4, 6]}$ (ATLAS), $S_5(B^0\to K^{*0}\mu^+\mu^-)^{[4, 6]}$ (ATLAS), $\frac{dB}{dq^2}(B^{+}\to K^{+}\mu^+\mu^-)^{[0.1, 0.98]}$ (LHCb) and $BR(B^{+}\to K^{+}\mu^+\mu^-)^{[1,6]}$ (Belle). Including all these, the list contains 12 data points which are explicitly shown in table \ref{tab:pullNP}.
\end{itemize}

For the observables listed in \textbf{List-1}, the respective values of the pull$^{NP}$ are given in table \ref{tab:pullNP}; these values are obtained for the one-operator scenario $\mathcal{O}_9$ with complex WCs. For the same scenario with real WC the respective pulls are also greater than 2. Following the discussions in items 2 and 3 above, we are not dropping the following data points from LHCb in the analyses: $P_5^{'\hspace{0.05 cm}[4, 6]}$, $S_5^{[4, 6]}$, $P_2^{[4, 6]}$, $R_{K^{*0}}^{[0.045,1.1]}$, $R_{K^{*0}}^{[1.1,6]}$ and $R_{K}^{[1.1,6]}$ despite their large pulls corresponding to the respective SM estimates.

\begin{table}[t]
	\small
	\begin{ruledtabular}
		\renewcommand*{\arraystretch}{1.2}
		\begin{tabular}{cccc}
			
			\multicolumn{4}{c|}{Fit scenario: Likelihood dataset 2020 ($\chi^2_{SM}= 288.9$,  $p$-value $(\%)$ in SM = 0.035)}    \\
			\hline
			$\chi _{Min}^2 / DOF $  &  $p$-value $(\%)$ &  Pull$_{SM}$ & confidence intervals \\
			\hline
			206.5/211 & 57.4 & 9.1 & $Re(\Delta C_9)\to -1.05 \pm 0.11 $  \\
			202.9/210 & 62.5 & 9.0 & $Re(\Delta C_9)\to -1.10 \pm 0.11$, $Im(\Delta C_9)\to 1.27^{+0.33}_{-0.43}$  \\
			\hline
			\multicolumn{4}{c}{Likelihood dataset 2020: without CP-asymmetric observables in $B_s \to \phi\mu\mu$ ($\chi^2_{SM}= 297.6$,  $p$-value $(\%)$ in SM = $9\times 10^{-4}$)  } \\
			\hline
			$\chi _{Min}^2 / DOF $ &  $p$-value $(\%)$  &  Pull$_{SM}$ & confidence intervals \\
			198.2/199 & 50.2 & 10 & $Re(\Delta C_9)\to -1.05 \pm 0.11 $ \\
			194.9/198 & 54.8 & 9.9 & $Re(\Delta C_9)\to -1.11^{+0.13}_{-0.12}$, $Im(\Delta C_9)\to \tcb{-}1.36^{+0.44}_{-0.34} \cup ~[0.84,~1.59]$ \\ 
			\hline
			\multicolumn{4}{c}{Fit scenario: Likelihood dataset 2020 $+$ CP-asymmetric observables in $B \to K^*\mu\mu$ from LHCb Likelihood dataset 2016}\\
			\multicolumn{4}{c}{ ($\chi^2_{SM}= 322.1$,  $p$-value $(\%)$ in SM = 0.09)}\\
			$\chi _{Min}^2 / DOF $ &  $p$-value $(\%)$  &  Pull$_{SM}$ & confidence intervals \\
			\hline
			239.8/246 &  60 & 9.1 & $Re(\Delta C_9)\to -1.06 \pm 0.11$ \\
			238.1/245 & 61.2 & 8.9 & $Re(\Delta C_9)\to -1.09 \pm 0.11$,~~~~$Im(\Delta C_9) \to -1.11^{+0.62}_{-0.40}$ 
		\end{tabular}
	\end{ruledtabular}
	\caption{\small Same as table \ref{tab:oneopralldata}, with the exception that now the observables given in \textbf{List-1}, has been dropped. The parameter uncertainties are estimated from profile likelihoods. }
	\label{tab:oneoprcwclis2}
\end{table}

We repeat all the fits with one operator after dropping these 12-data points from our complete list. After dropping the observables given in \textbf{List-1}, we have noticed a considerable improvement in the overall fit quality. Other than $\mathcal{O}_9$, the fit-qualities for all the other one-operator scenarios, though better than those before dropping these data, are still very poor. For $\mathcal{O}_9$, we have presented the corresponding results in table \ref{tab:oneoprcwclis2}. The respective improvements in the fit scenarios compared to those given in table \ref{tab:oneopralldata} are apparent. Note that the allowed parameter spaces for $\Delta C_9$ are consistent in the fits with and without the data dropped from \textbf{List-1}, respectively. We observe a slight improvement in the fit-quality for the analysis with complex WCs compared to real WCs.

 For the current datasets, we do not have many references to cross-check our findings. At the moment, we are aware of a few analyses which have used 2020 LHCb datasets \cite{Alguero:2019ptt, Hurth:2020rzx,Altmannshofer:2021qrr,Carvunis:2021jga}. However, none of these considers the exact same dataset as ours; only a subset has been used. At least in \cite{Altmannshofer:2021qrr,Carvunis:2021jga}, the authors perform a global fit of complex $C_9$ WC including LHCb 2020 data on $B^{0}\to K^{*0}\mu^+\mu^-$ \cite{Aaij:2020nrf}. In ref. \cite{Hurth:2020rzx}, the authors have performed the analysis with 117 data points which include a considerable number of data from $\Lambda_b\to \Lambda \mu\mu$. The analysis of ref. \cite{Alguero:2019ptt} finds the above scenario to be favourable. However, in \cite{Alguero:2019ptt}, the authors have considered a total of 180 data points. Although many of them coincide with ours, there are appreciable differences. For example, they have not considered any CP-averaged observables ($S_i$), forward-backward asymmetries $A_{FB}$ from LHCb, measured values of branching fractions and isospin asymmetries from Belle, a few optimized observables in large bins from LHCb, asymmetric observables in $B_s\to \phi\mu\mu$ decays etc. Also, they have considered a few data points corresponding to the high-$q^2$ regions, which we don't consider. Therefore, in general, it would be hard to compare our findings with them directly. However, we would like to mention that out of the 12 data points listed in table \ref{tab:pullNP} only 4 data points are considered in the analysis of \cite{Alguero:2019ptt}. They have included the following four data points: $P_4^{'}(B^0\to K^{*0}\mu^+\mu^-)^{[4,6]}$ (ATLAS), $P_6^{'}(B^0\to K^{*0}\mu^+\mu^-)^{[4,6]}$ (LHCb) and $P_5^{'}(B^0\to K^{*0}\mu^+\mu^-)^{[4, 6]}$ (ATLAS), $\frac{dB}{dq^2}(B^{+}\to K^{+}\mu^+\mu^-)^{[0.1, 0.98]}$ (LHCb), rest of the 8 data points are not considered. 
 
 A few additional remarks on the impact of the data points listed in table \ref{tab:pullNP} on the fit to `Likelihood dataset 2020':
\begin{itemize}
	\item Note that the data points listed in tables \ref{tab:pullsliklihood} and \ref{tab:pullcommon} with higher pulls (w.r.t. the SM values) also have higher values of pull$^{NP}$ obtained using eq. \ref{eq:pullNP}.
	
	\item In the analysis of \textbf{Likelihood dataset 2020}, if we drop only the observables listed in the left column of table \ref{tab:pullNP}, we  get an allowed fit for the scenario $\mathcal{O}_9$ with a p-value $\approx$ 7\%. 	 
	
	\item We have noted that the inclusion of data points: $P_4^{'}(B^0\to K^{*0}\mu^+\mu^-)^{[4,6]}$ (ATLAS), $P_6^{'\hspace{0.05 cm}[4, 6]}$ (LHCb) and $P_5^{'}(B^0\to K^{*0}\mu^+\mu^-)^{[4, 6]}$ (ATLAS) from \textbf{List-1} reduce the fit-quality from 62.5\% to 36\%. In addition, if we include $\frac{dB}{dq^2}(B^{+}\to K^{+}\mu^+\mu^-)^{[0.1, 0.98]}$ (LHCb), the $p$-value of the fit will reduce further by a few percent. Hence, we can get an allowed fit with an appreciable p-value for the scenario $\mathcal{O}_9$ even if we include these three inputs on optimized observables.
	  
\end{itemize}

\begin{figure*}[t]
	\small
	\centering
	\subfloat[]{\includegraphics[width=0.5\textwidth]{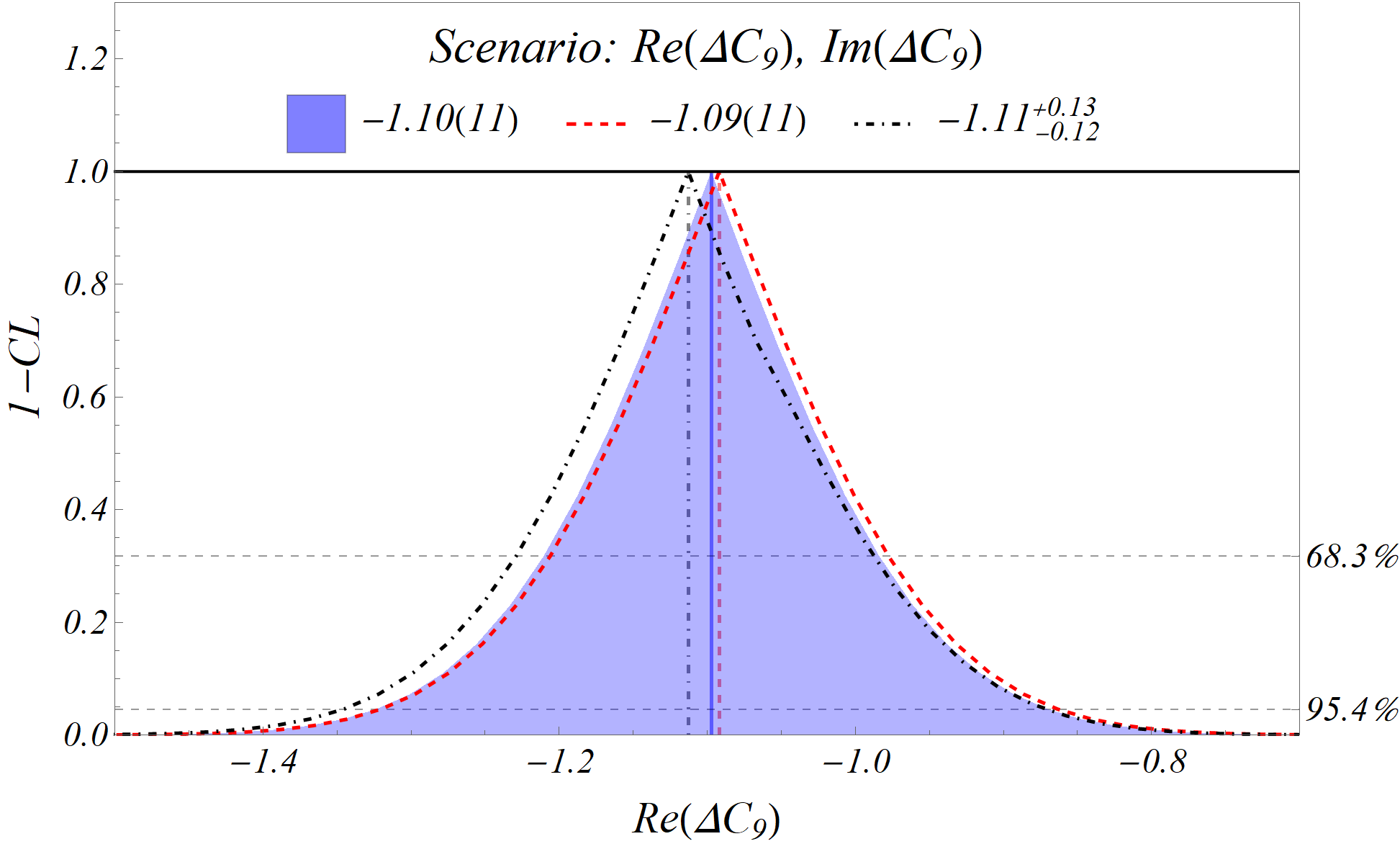}\label{fig:cwcrec9}}~
	\subfloat[]{\includegraphics[width=0.5\textwidth]{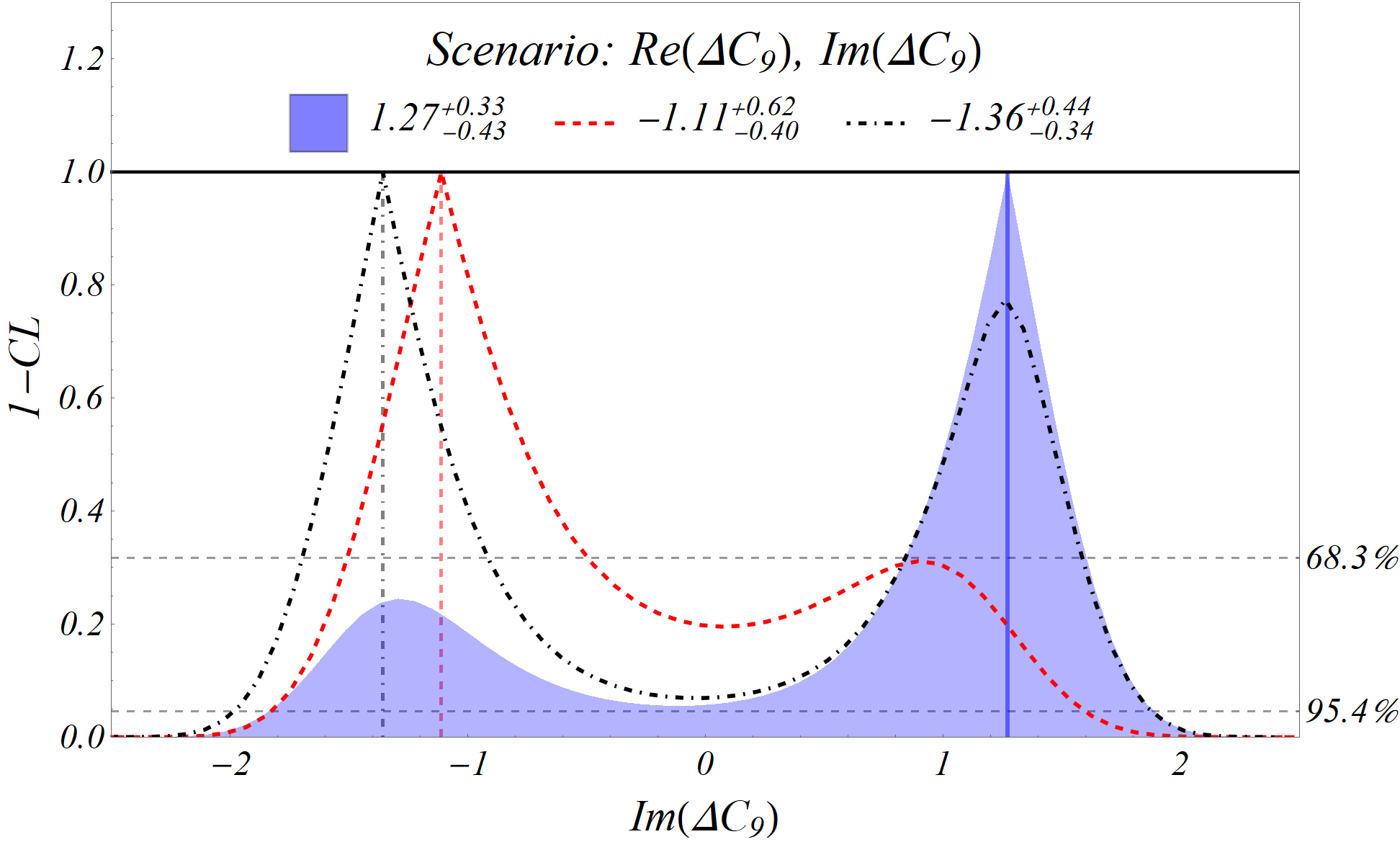}\label{fig:imc9}}\\
	\subfloat[]{\includegraphics[width=0.5\textwidth]{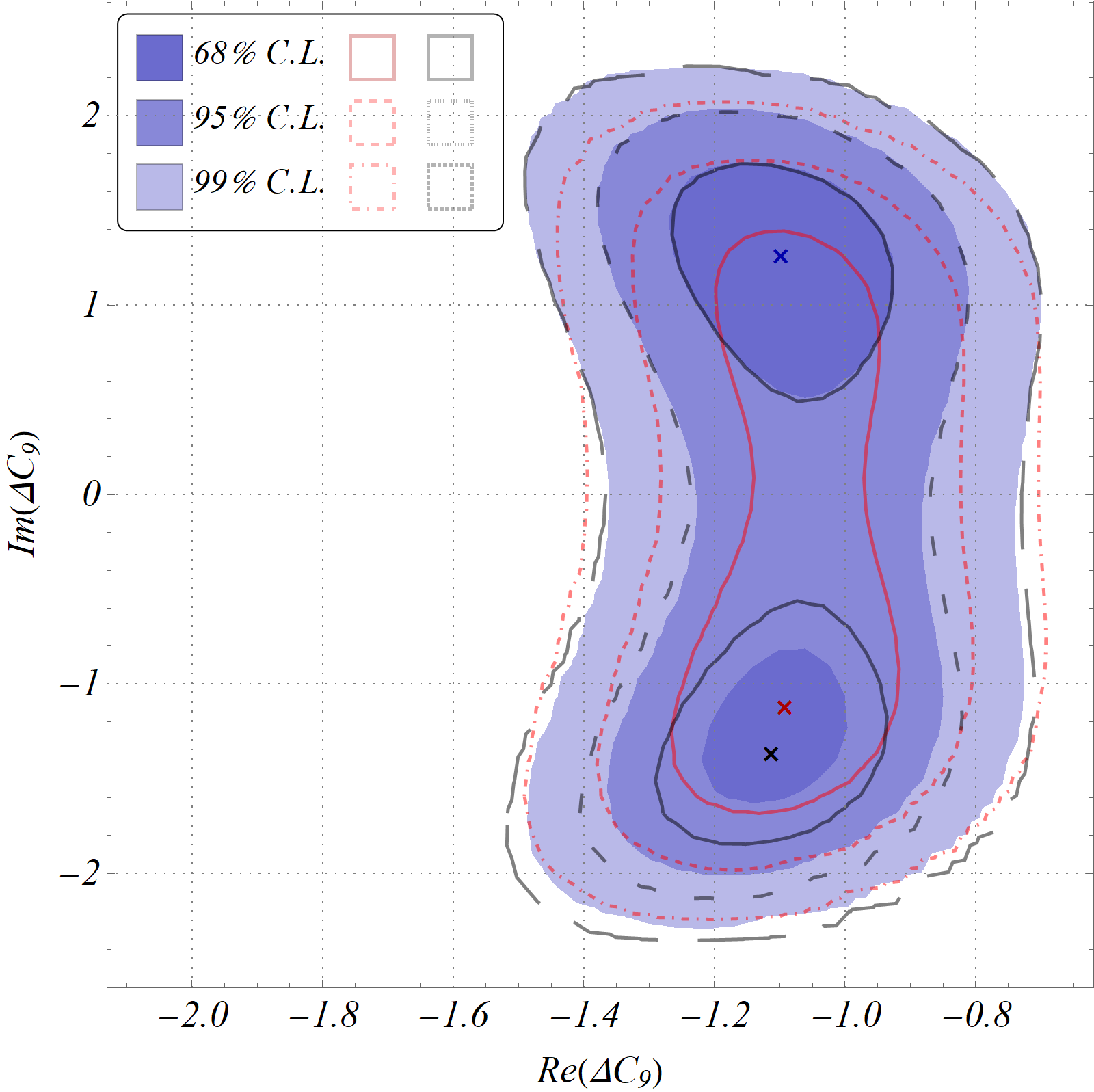}\label{fig:corr}}~
	\subfloat[]{\includegraphics[width=0.4\textwidth]{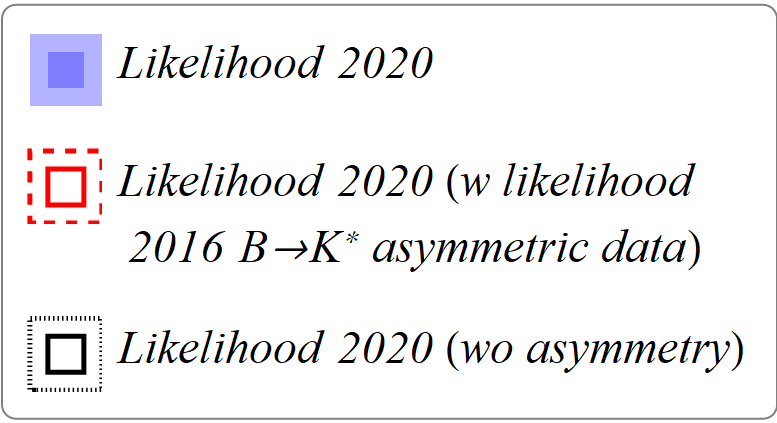}\label{fig:legend}}
	\caption{One and two-parameter profile-likelihoods corresponding to the single operator scenario $O_9$ with complex WC in different fit scenarios as discussed in table \ref{tab:oneoprcwclis2}. (a) and (b) display the one parameter profile likelihoods for the real and imaginary parts of $\Delta C_9$ while fig. (c) displays the two parameter profile likelihood. The filled, blue contour corresponds to the new ``Likelihood 2020 datasets" and the red dashed contour to the same datasets including the CP-asymmetric observables in $B \to K^*$ from ``Likelihood 2016 datasets", respectively. The black, dot-dashed contour represents the ``Likelihood 2020" data with all the asymmetric observables (due to $B_s\to\phi\mu\mu$ modes) removed. The corresponding legends are shown in figure (d).}
	\label{fig:param}
\end{figure*}

Estimation of the parameter-uncertainties from the Hessian matrix of the $\chi^2$ assumes Gaussian nature for the said uncertainties, and is not always accurate. On the other hand, confidence intervals estimated from the actual 1-D profile likelihoods correctly capture the non-uniformity of the uncertainties (also, ambiguities), and thus are considered to be more accurate. Following this reasoning, we have chosen to showcase our results in the latter form in table  \ref{tab:oneoprcwclis2}. The corresponding 1-D and 2-D profile likelihoods for the real and imaginary WCs have been depicted as 1-CL plots, which are shown in fig.~\ref{fig:param}. In 2-D, $x~\sigma$ CL corresponds to the region whose projection on both axes would bound $x~\sigma$ CL for the corresponding 1-D parameter space. For example, in 2-D, $1~\sigma \equiv 39.347\%$CL.

\begin{figure*}[t]
	\small
	\centering
	\subfloat[]{\includegraphics[width=0.3\textwidth]{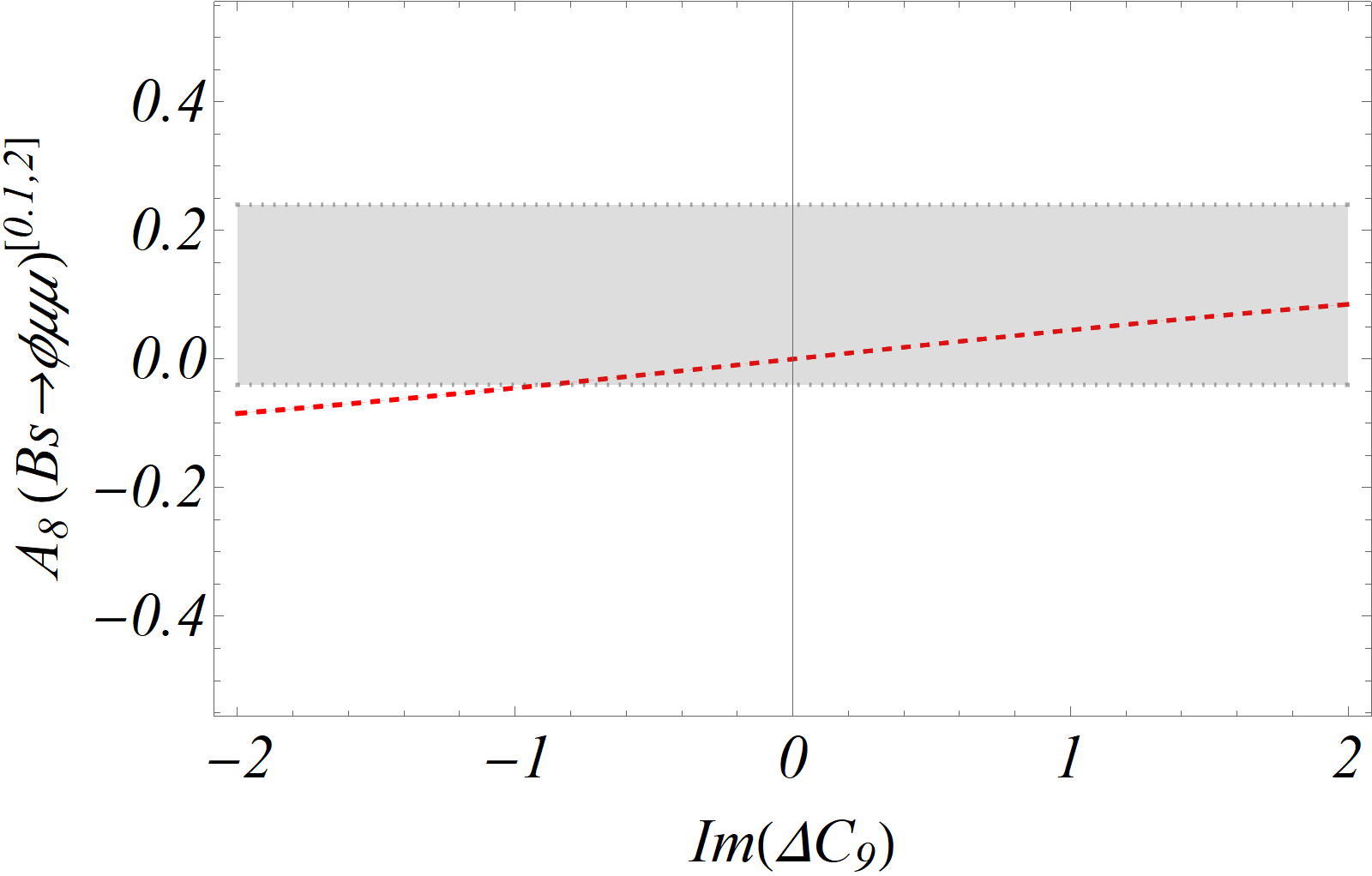}\label{fig:A801to2}}~~~
	\subfloat[]{\includegraphics[width=0.3\textwidth]{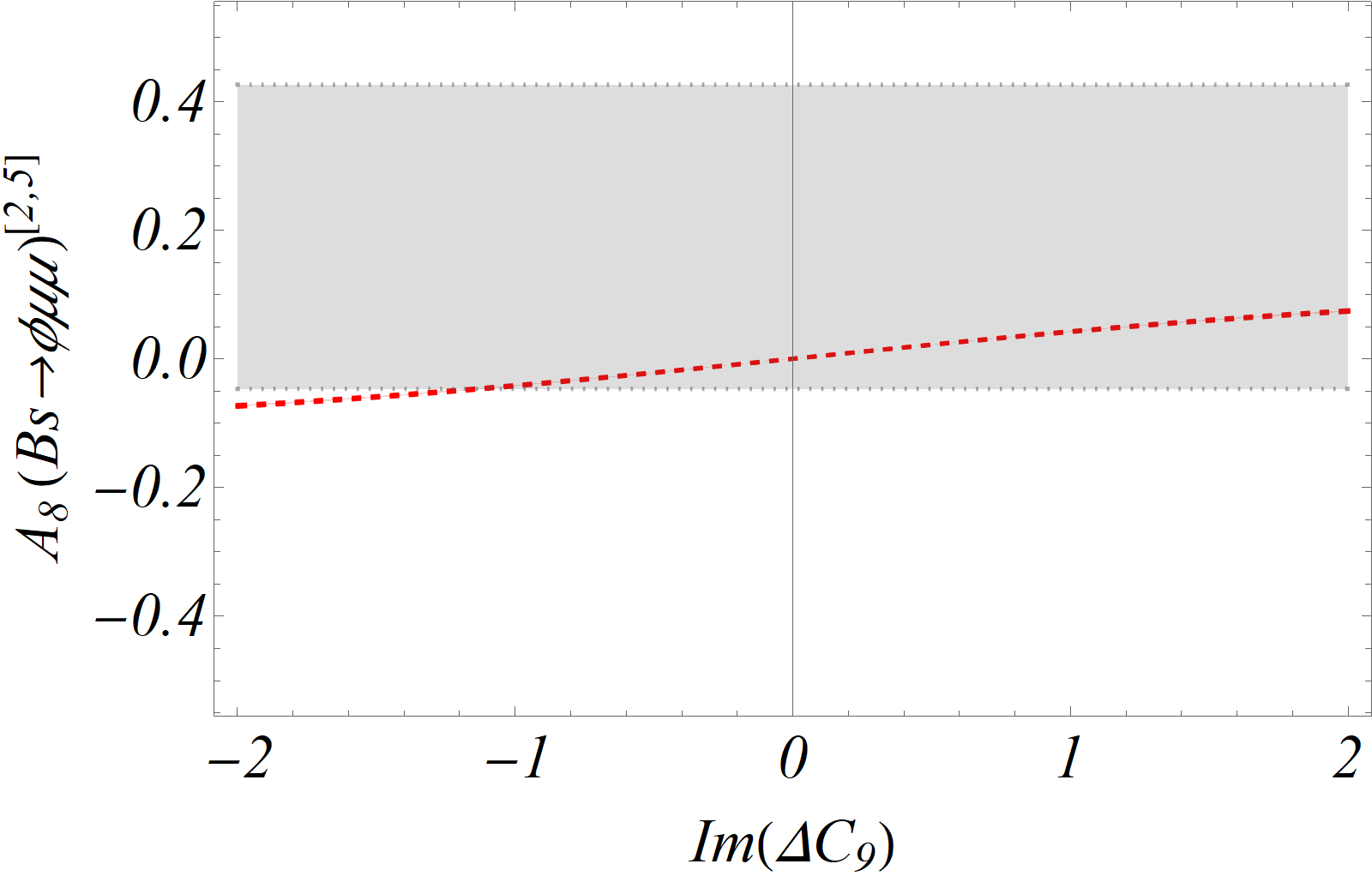}\label{fig:A82to5}}~~~
	\subfloat[]{\includegraphics[width=0.3\textwidth]{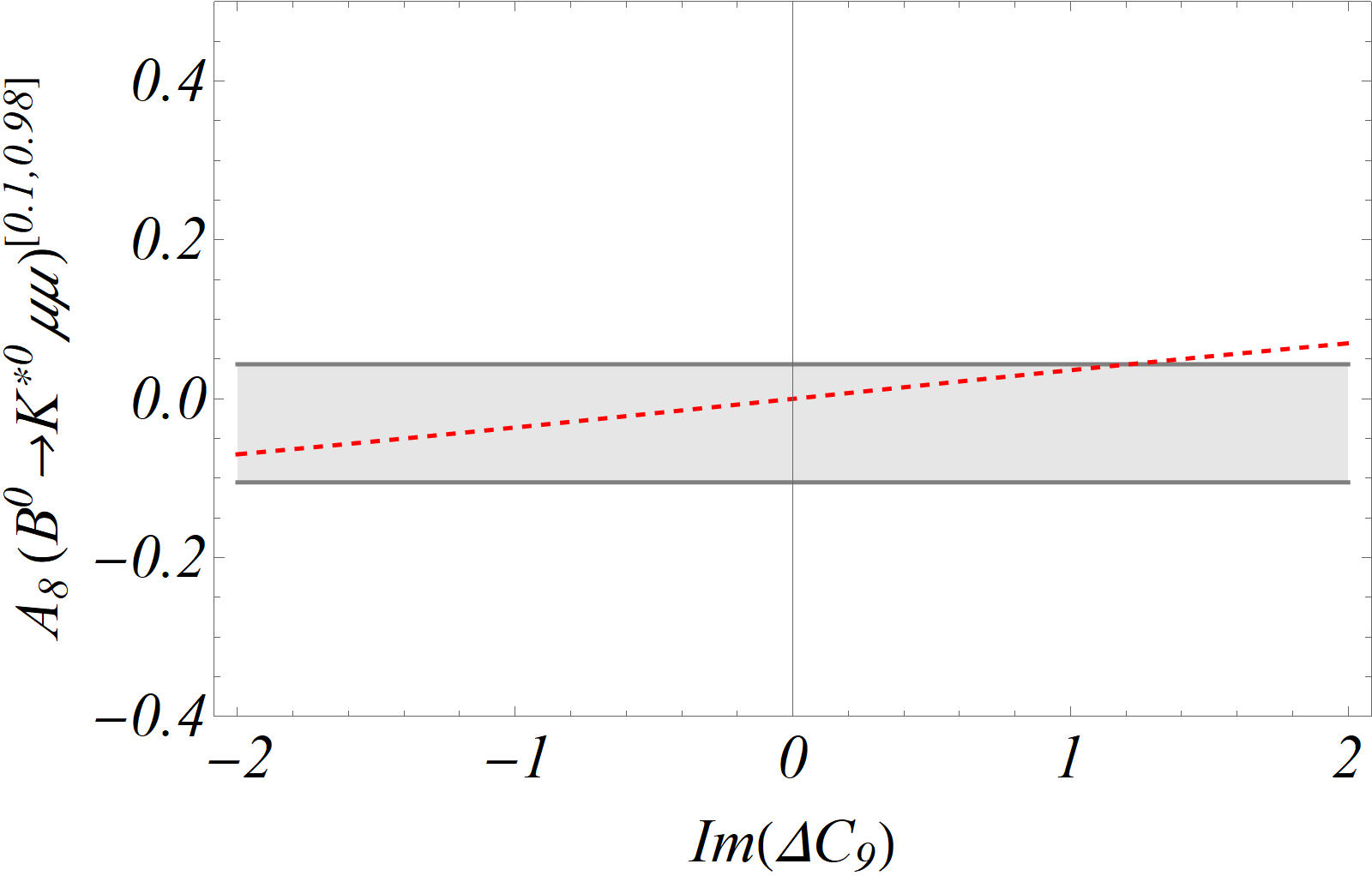}\label{fig:A801098}}\\
	\subfloat[]{\includegraphics[width=0.3\textwidth]{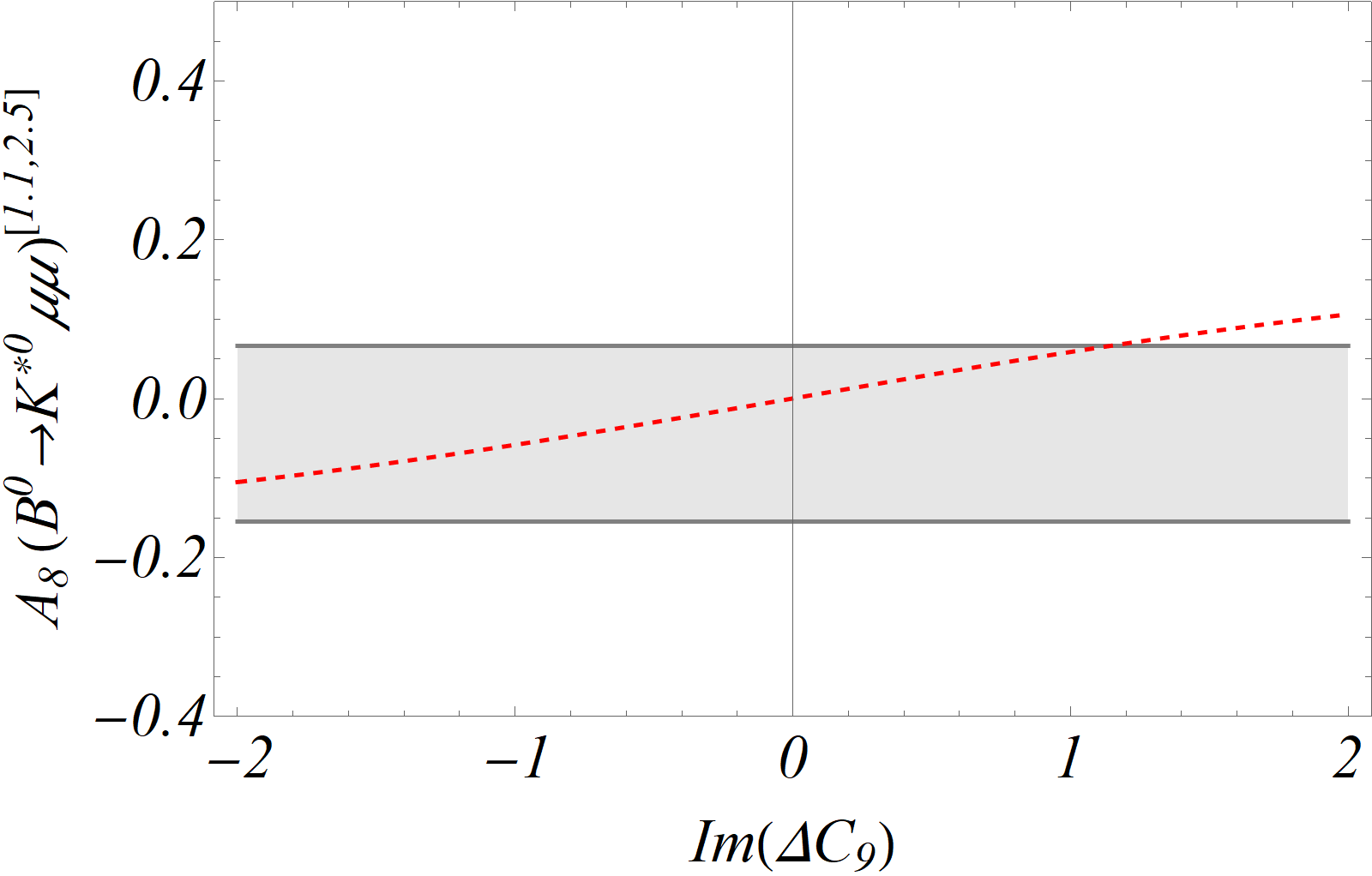}\label{fig:A81125}}~~~
	\subfloat[]{\includegraphics[width=0.3\textwidth]{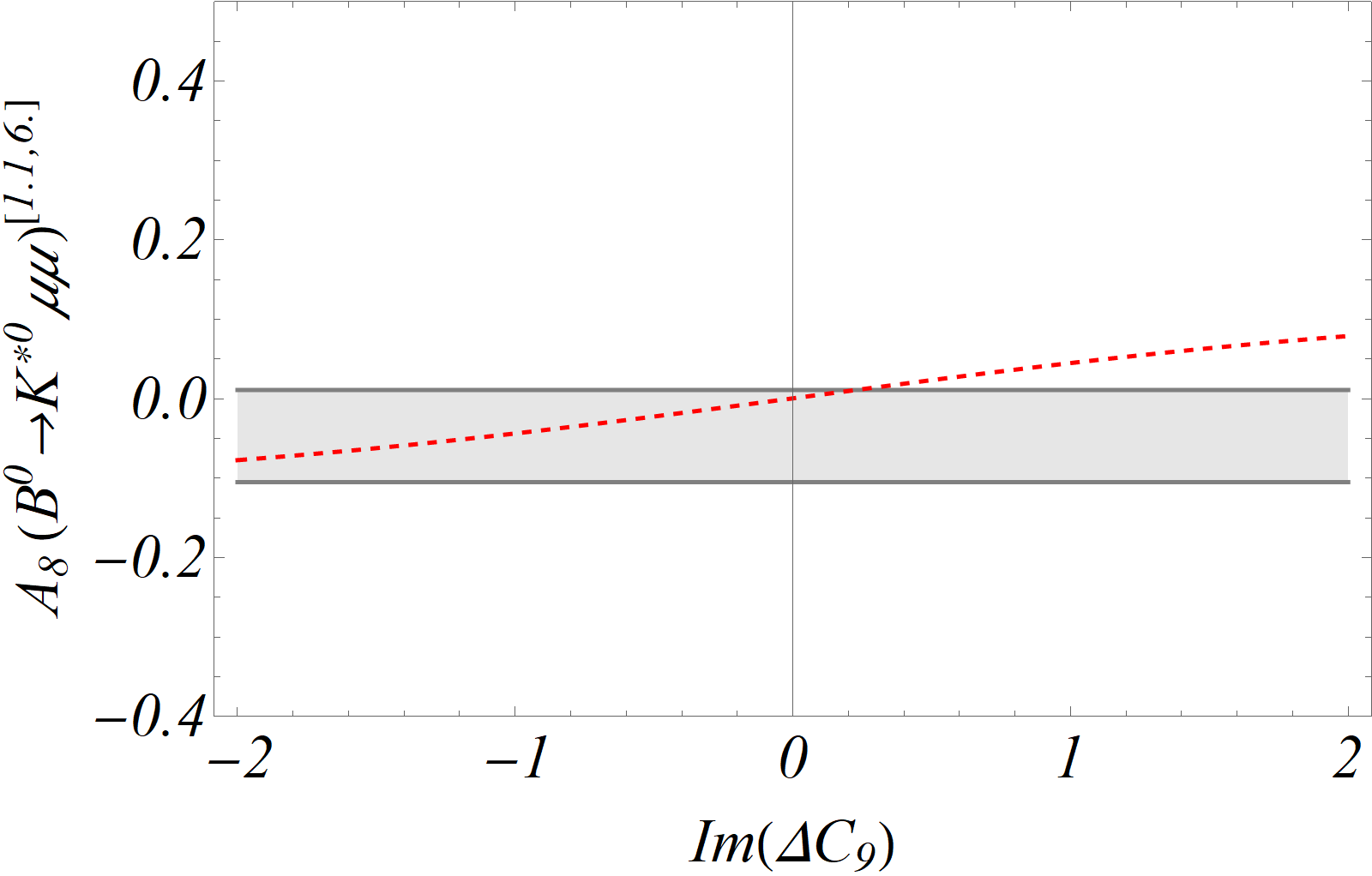}\label{fig:A8116}}~~~
	\subfloat[]{\includegraphics[width=0.3\textwidth]{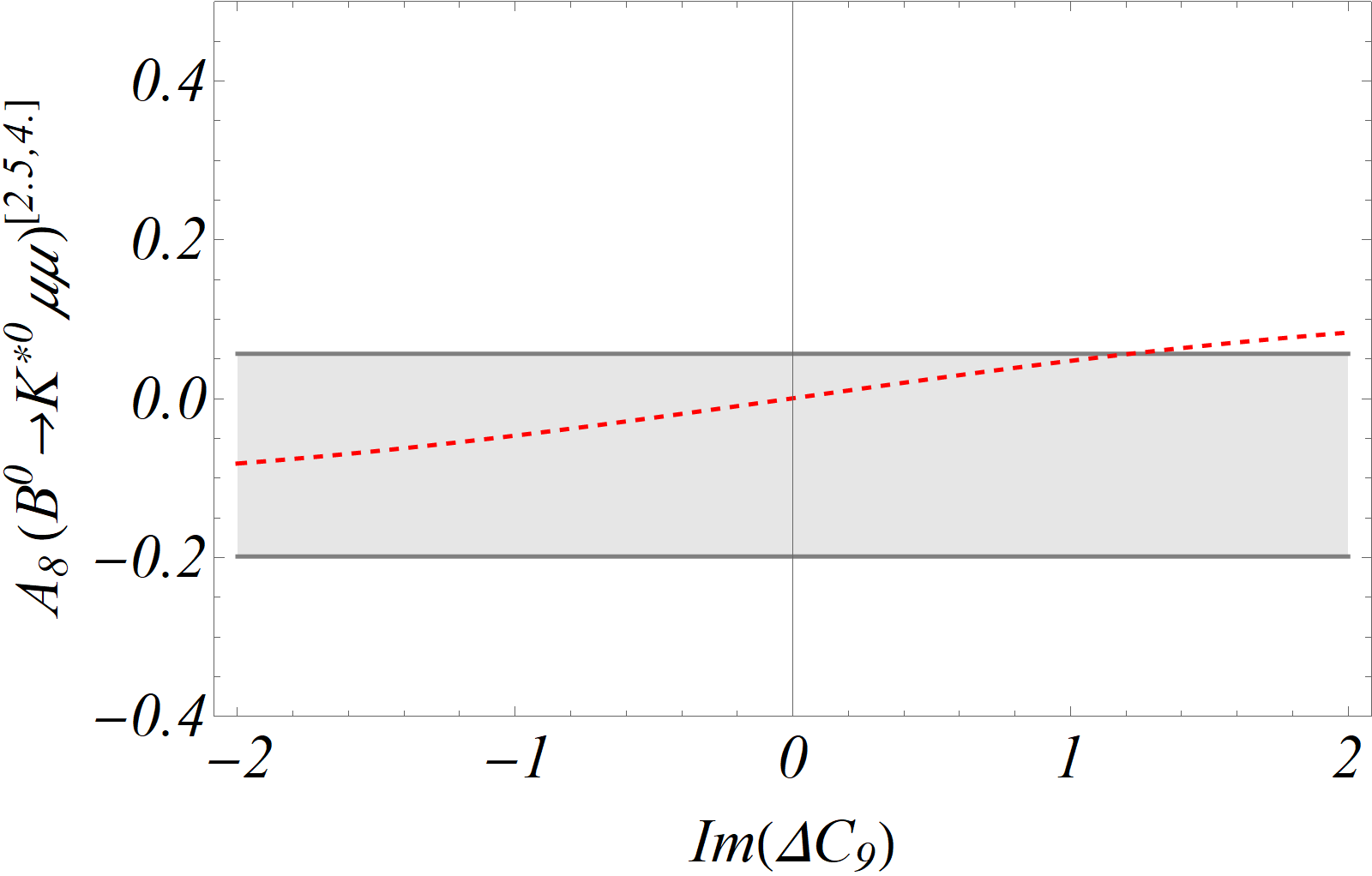}\label{fig:A8254}}
	\caption{ Sensitivity of a few of CP asymmetric observables in $B_s \to\phi\mu\mu$ and $B \to K^*$ decays to the Imaginary part of $\Delta C_9$. The grey bands represent the current LHCb bounds.}
	\label{fig:imgsenbstophi}
\end{figure*}

\begin{figure*}[htbp]
	\small
	\centering
	\subfloat[]{\includegraphics[width=0.3\textwidth]{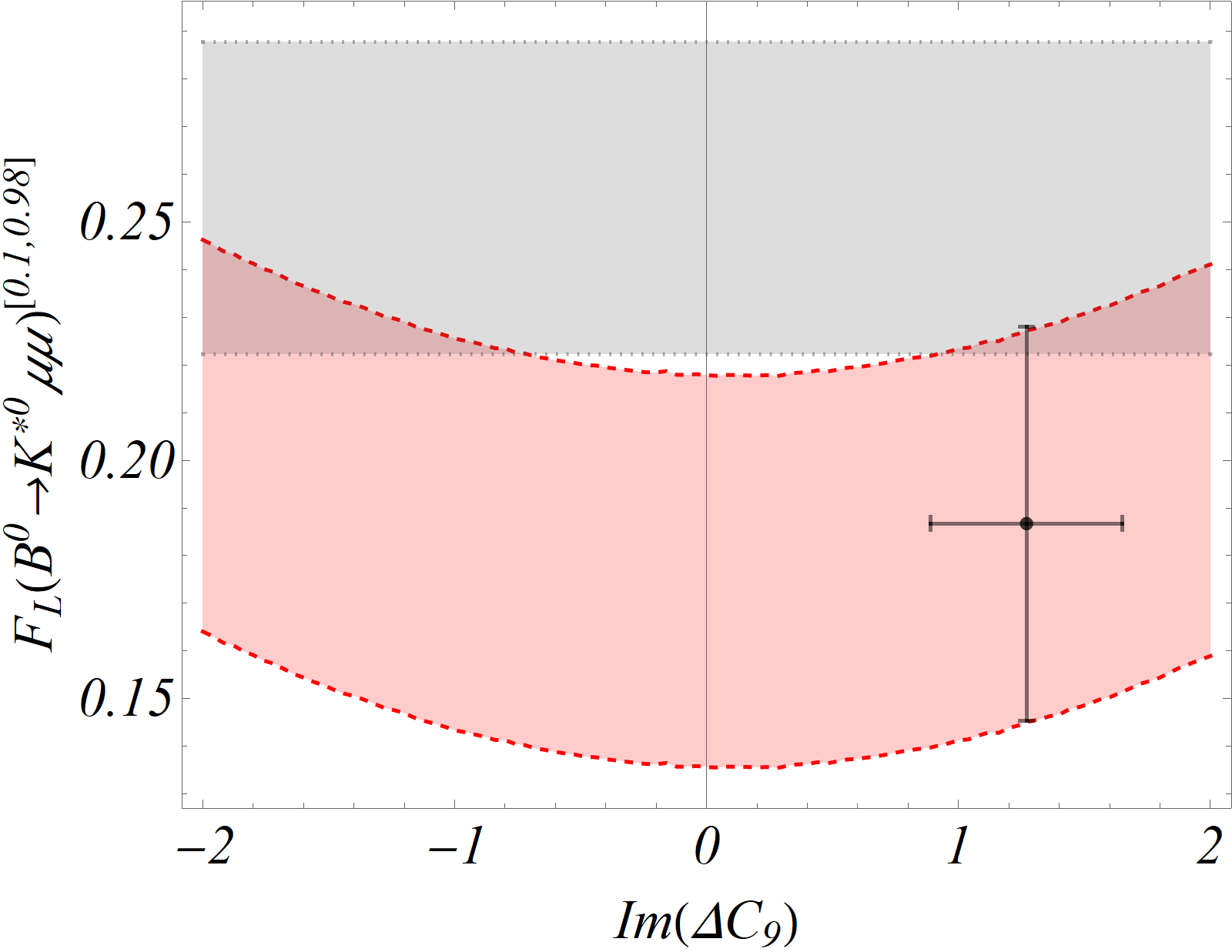}\label{fig:fl}}~~~
	\subfloat[]{\includegraphics[width=0.3\textwidth]{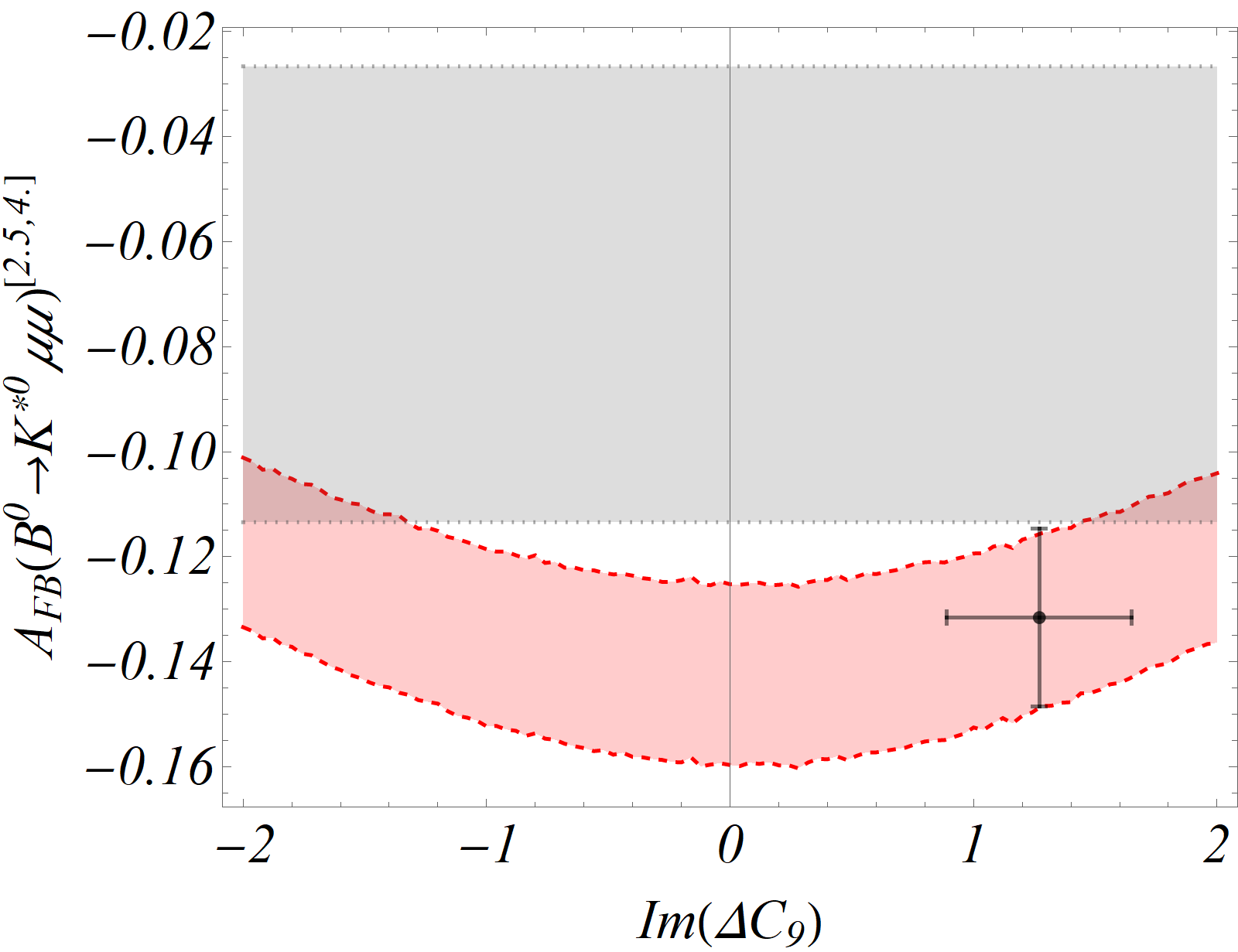}\label{fig:afb}}~~~
	\subfloat[]{\includegraphics[width=0.3\textwidth]{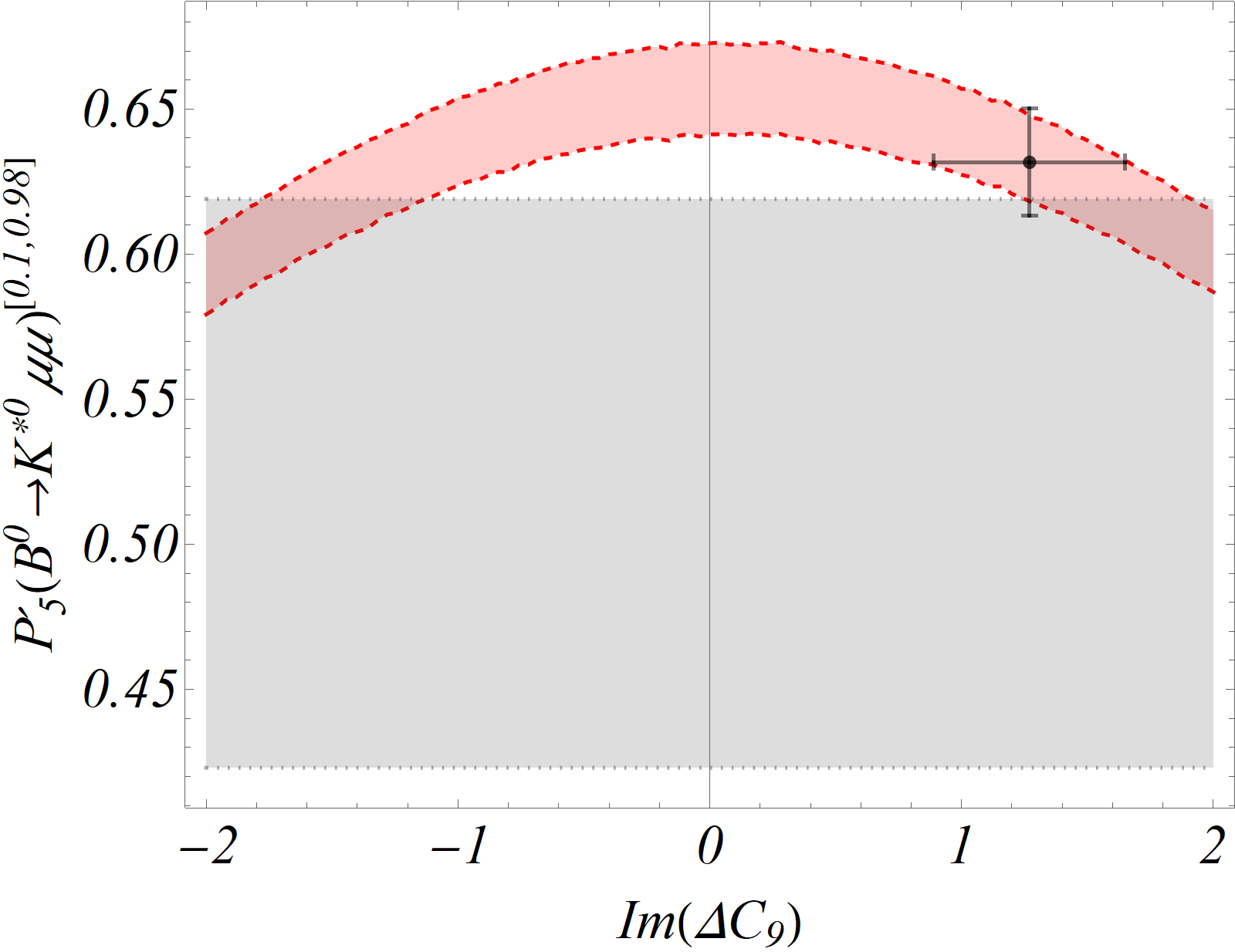}\label{fig:p5pr}}\\
	\subfloat[]{\includegraphics[width=0.3\textwidth]{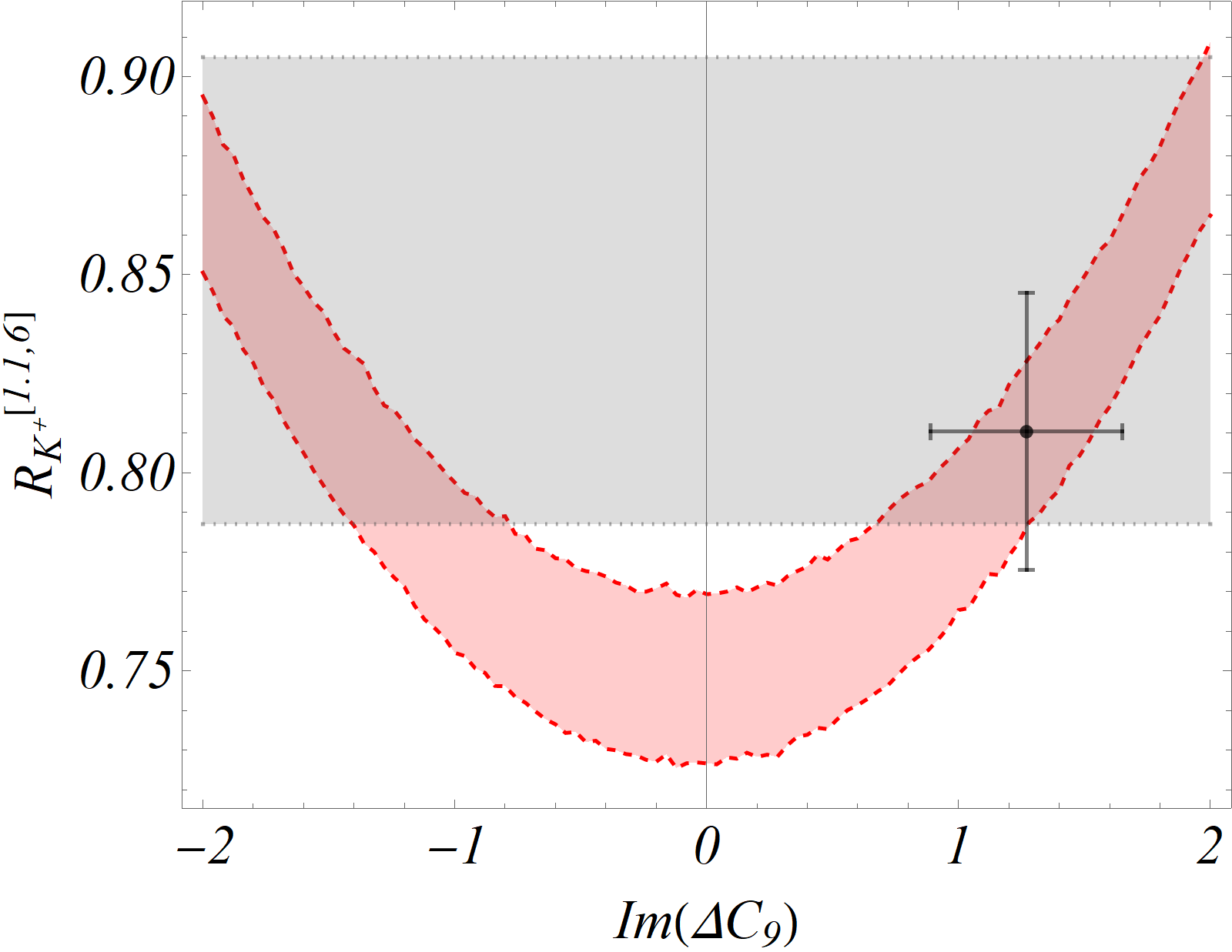}\label{fig:rkp}}~~~
	\subfloat[]{\includegraphics[scale=0.6]{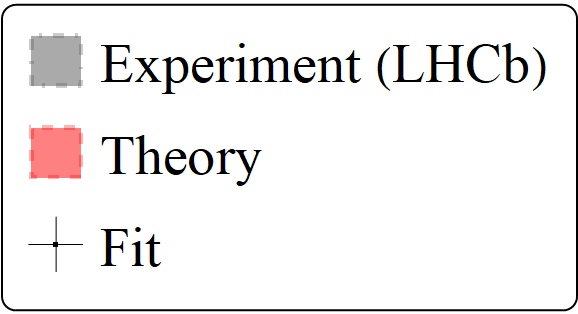}}
	\caption{ Sensitivity of a few of CP symmetric observables to the Imaginary part of $\Delta C_9$. }
	\label{fig:imgsensitivity}
\end{figure*}

As mentioned earlier, in figs. \ref{fig:cwcrec9} and \ref{fig:imc9}, we have compared the 1-D profile likelihoods of $Re(\Delta C_9)$ and $Im(\Delta C_9)$ obtained from fits to the different datasets as given in table \ref{tab:oneoprcwclis2}. In fig.\ref{fig:corr}, we have shown the corresponding 2-D profile likelihoods in the $Re(\Delta C_9)- Im(\Delta C_9)$ plane. Following are the observations made from these profile likelihoods: 
\begin{itemize}
	\item As shown in the table \ref{tab:oneoprcwclis2}, the allowed confidence intervals for $Re(\Delta C_9)$ are consistent in all the three fit scenarios.
	
	\item At $68.3\% CL$, we note a significant shift in the allowed regions of $Im(\Delta C_9)$ in the fits with and without the CP-asymmetric observables in $B\to K^*$ modes, and they are almost mirror images of each other.
	
	\item The fit to the datasets without the CP-asymmetric observables in $B\to K^*$ modes has a slight preference for positive solutions for $Im(\Delta C_9)$, though negative solutions are allowed. Also at $95\% CL$, the allowed regions are consistent with zero (see figure \ref{fig:corr}).  
	
	\item On the other hand, the fit to the datasets with all the CP-asymmetric observables in $B\to K^*$ and $B_s\to \phi$ modes prefers negative solutions for $Im(\Delta C_9)$, though positive solutions are also allowed. In this case, even at 68\% CL the allowed regions are consistent with zero (see figure \ref{fig:corr}).{\tiny }
	
	\item In the fit with all the CP-asymmetric observables dropped, both positive and negative solutions for $Im(\Delta C_9)$ are allowed and both of them can be considered favorable. As can be seen from fig. \ref{fig:corr}, at $68.3\% CL$ we have two disconnected regions that are almost symmetric w.r.t zero; while at $95\% CL$ we get a zero consistent solution. In particular, we have checked that the $\chi^2_{min}$ for the negative and positive solutions are $194.93$ and $195.03$, respectively\footnote{A frequentist chi-square minimization test of the data will converge only to the negative solution since for this solution the $\chi^2$ is minimum, though the value of $\chi^2_{min}$ for the positive solution is only slightly greater. To get the 2-D profile likelihood, we need to scan over the parameter spaces for the real and imaginary $\Delta C_9$. If we do not scan over a broad range, then there is a possibility of missing the other solution. In our case, we scan over $-2.5 \le Im(\Delta C_9) \le 2.5 $.}. 
	
	\item  The allowed ranges of $Im(\Delta C_9)$ will be sensitive to the choices of the CKM elements since in the SM, the complex part of $C_9$ is sensitive to the CKM element $V_{ub}$ which has a phase. In this analysis, we have presented our results using the inputs from CKMfitter group \cite{CKMfitter} where the analysis has been done assuming SM for the $\Delta F=2$ processes\footnote{The UTfitter group has presented their results assuming SM, as well as considering NP effects in $\Delta F =2$ processes \cite{UTfitter}. There are small differences between the two sets of results. We have checked that if we incorporate the inputs of the fit with NP, the best fit value of $Im(\Delta C_9)$ will be higher by $\approx 5\%$, though they are consistent within the respective error bars.}.    	
\end{itemize}

It is possible to understand these observations from a study of the dependencies of the CP-asymmetric and other associated angular observables in $B\to K^*$ and $B_s\to \phi$ decay modes on $Im(\Delta C_9)$. We have figured out a few CP-asymmetric observables in $B_s\to \phi\mu\mu$ decays, for example, $A_8^{[0.1,2]}$ and $A_8^{[2,5]}$ for which, at the present level of accuracy, the current measured values prefer the positive values of $Im(\Delta C_9)$ slightly. The dependence of these observables on $Im(\Delta C_9)$ for the best fit value of $Re(\Delta C_9)$ are shown in figure \ref{fig:imgsenbstophi}. Note that though these observables prefer a positive value for $Im(\Delta C_9)$, negative values are also allowed. This is also evident from the corresponding profile likelihood as discussed above. 

On the other hand, from the CP-asymmetric observables in $B\to K^*$ modes we have found that $A_8^{[0.1,0.98]}$, $A_8^{[1.1,2.5]}$, $A_8^{[1.1,6]}$ and $A_8^{[2.5,4]}$ favor negative values of $Im(\Delta C_9)$, which are shown in fig. \ref{fig:imgsenbstophi}, though positive values are also allowed. In general, for both $B\to K^*$ and $B_s\to \phi$ modes the dependence of $A_8$ on $Im(\Delta C_9)$ should be similar. However, as can be seen from the figures, the deciding factors in this case are the respective measured values. Note that the measured values of $A_8$ in $B_s \to \phi$ have large errors and are slightly more positive than the ones measured in $B \to K^*$ modes. The data for the $B\to K^*$ modes being relatively more precise, they play the dominant role in choosing the favorable solutions for $Im(\Delta C_9)$ when combined with the CP-asymmetric observables due to $B_s\to \phi$.    

In all the three fit scenarios, with and without the CP-asymmetric observables, a considerable value of $Im(\Delta C_9)$ is allowed. Note that large values of $Im(\Delta C_9)$ are allowed even when the CP-asymmetric observables in $B\to K^*$ and $B_s\to \phi$ decays are not included in the fit, though at $95\% CL$ the solutions are consistent with zero. The requirement of a large $Im(\Delta C_9)$, even in the absence of any CP-asymmetric observables can be understood by looking at the dependence of other observables on $Im(\Delta C_9)$. We have pointed out a few such observables in this work. In figs.~\ref{fig:fl}, \ref{fig:afb}, \ref{fig:p5pr} and \ref{fig:rkp}, we have shown the variation of $F_L$, $A_{FB}$, $P_5^{'}$ and $R_{K^+}$ respectively with $Im(\Delta C_9)$ for the allowed values of $Re(\Delta C_9)$ (at $68\% CL$), as given in Table \ref{tab:oneoprcwclis2}. We display only those $q^2$-bins that exhibit the possibility of a large non-zero imaginary contribution from $\Delta C_9$. As can be seen from the figures, we require a large non-zero $Im(\Delta C_9)$ to explain the experimental observations within their 1-$\sigma$ confidence interval in a few specified $q^2$-bins. Moreover, the allowed values are symmetric about $Im(\Delta C_9) = 0$. Though it is too early to conclude, at the moment it might be hinting towards the possibility of a large imaginary contribution to $\Delta C_9$. However, further assertions and hence solid conclusions must be subject to more precise data, hopefully in the near future.

\begin{table}[t]
	\begin{ruledtabular}
		\renewcommand*{\arraystretch}{1.2}
		\begin{tabular}{|c|c|cccc|}
			\multicolumn{6}{|c|}{The new physics scenario $\Delta C_9 = -\Delta C_{10}$}\\
			\hline
			Fit scenario & Data dropped &$\chi _{Min}^2 / DOF $ & $p$-value $(\%)$ & Pull$_{SM}$ & Confidence intervals \\
			\hline
			& \textbf{List-1} & 243.02/211 & 6.4 & 6.8 & $Re(\Delta C_9) = - Re(\Delta C_{10}) \to - 0.55 \pm 0.08$  \\
			\multirow{6}{*}{Liklihood datasets 2020}& $\chi^2_{SM}= 288.9$   & 242.58/210 & 6.1 & 6.5 & $Re(\Delta C_9) = - Re(\Delta C_{10}) \to - 0.59 \pm 0.12$  \\
			&  $p_{SM}$ = 0.035 $\%$ & & & & $Im(\Delta C_9) = - Im(\Delta C_{10}) \to 0.45 \pm 0.49$ \\
			\cline{2-6}
			& \textbf{List-2} & 209.8/204 & 37.6 & 6.8 & $Re(\Delta C_9) = -Re(\Delta C_{10})\to -0.54 \pm 0.08$\\
			& $\chi^2_{SM}= 255.4$ & 209.6/203 & 36 & 6.4 & $Re(\Delta C_9) = - Re(\Delta C_{10})\to -0.56 \pm 0.10$\\
			& $p_{SM}$ = 0.96 $\%$ & & & & $Im(\Delta C_9) = - Im(\Delta C_{10}) \to 0.27 \pm 0.56$\\                  
			\hline	
			\multirow{4}{*}{Liklihood datasets 2016} & \textbf{List-3}  & 269.15/250 &  19.3 & 6.5 & $Re(\Delta C_9) = - Re(\Delta C_{10}) \to - 0.54 \pm 0.09$  \\
			& $\chi^2_{SM}= 311.5$ & 268.84/249 & 18.5 & 6.2 & $Re(\Delta C_9) = - Re(\Delta C_{10}) \to - 0.54\pm 0.09$  \\
			& $p_{SM}$ = 0.56 $\%$ & & & & $Im(\Delta C_9) = - Im(\Delta C_{10}) \to -0.16 \pm 0.28$ \\
			\hline		
			\multirow{4}{*}{Moment datasets 2016} & \textbf{List-3} & 260.82/272 &  67.6 & 5.9 & $Re(\Delta C_9) = - Re(\Delta C_{10}) \to - 0.49 \pm 0.09$  \\
			&  $\chi^2_{SM}= 295.6$  & 260.72/271 & 66.2 & 5.6 & $Re(\Delta C_9) = - Re(\Delta C_{10}) \to - 0.49\pm 0.09$  \\
			& $p_{SM}$ = 16.6 $\%$ & & & & $Im(\Delta C_9) = - Im(\Delta C_{10}) \to 0.14 \pm 0.43$ \\
		\end{tabular}
	\end{ruledtabular}
	\caption{The case study of new physics scenario $\Delta C_9 = -\Delta C_{10}$ in the couple of different fit procedure.}
	\label{tab:c9eqmc10}
\end{table}

In passing, we also carry out the fit for the interesting scenario $\Delta C_9 = - \Delta C_{10}$. The analysis with the LHCb `Moment datasets' shows this scenario to be a favorable solution for the data, which is in agreement with earlier observations (see, for example, \cite{Altmannshofer:2014rta,Geng:2017svp}). Also, for the `Liklihood 2020 datasets', this scenario has been shown as a favourable one in \cite{Alguero:2019ptt}. For the NP scenario as mentioned above, we have done a comparative study in all the different datasets given earlier. At first, we have carried out the fit to this scenario after dropping the inputs given in 'List-1'. The corresponding result is shown in the first row in Table \ref{tab:c9eqmc10}. The fit has a $p$-value $\sim$ 6\%, which can be considered as a marginally significant result. Therefore, to look for a possibility of improvement in the quality of fit, we have prepared a second list with the following observables, 
 \begin{itemize}	
  	\item \textbf{List-2}: It contains all the observables in table \ref{tab:pullcommon} and the observables from LHCb likelihood 2020 dataset in table \ref{tab:pullsliklihood} with a pull $>$ 2 excluding the following: $A_{FB}^{[0.1,0.98]}$, $P_5^{'\hspace{0.05 cm}[4, 6]}$, $S_5^{[4, 6]}$, and $P_2^{[4, 6]}$\footnote{Note that the only available data on $A_{FB}^{[0.1,0.98]}$ is from LHCb; therefore, we keep it in our analysis, though these measurements have large errors and a pull $>$ 2. Following the earlier discussion, we don't drop $P_5^{'\hspace{0.05 cm}[4, 6]}$, $S_5^{[4, 6]}$, and $P_2^{[4, 6]}$ from the analysis}. On top of this, we have dropped few more observables like $P_5^{'}(B^0\to K^{*0}\mu^+\mu^-)^{[4, 6]}$ (ATLAS), $S_5(B^0\to K^{*0}\mu^+\mu^-)^{[4, 6]}$ (ATLAS), $\frac{dB}{dq^2}(B^{+}\to K^{+}\mu^+\mu^-)^{[0.1, 0.98]}$ (LHCb) and $BR(B^{+}\to K^{+}\mu^+\mu^-)^{[1,6]}$ (Belle). Overall, this list contains 19 data points.
\end{itemize}

\begin{table}[htbp]
	\begin{center}
		\begin{tabular}{|c|}
			\hline
			\multicolumn{1}{|c|}{Additional observables in \textbf{List-2} with the respective pulls in the one-operator scenario $\mathcal{O}_9$}  \\
			\hline  
			$S_3(B^0\to K^{*0}\mu^+\mu^-)^{[1.1, 2.5]}$ (LHCb) $\rightarrow$ - 2.06  \\
			$S_8(B^0\to K^{*0}\mu^+\mu^-)^{[1.1, 2.5]}$ (LHCb) $\rightarrow$ - 2.37 \\
			$S_9(B^0\to K^{*0}\mu^+\mu^-)^{[1.1, 2.5]}$ (LHCb) $\rightarrow$ - 2.07\\
		   	$P_1(B^0\to K^{*0}\mu^+\mu^-)^{[1.1, 2.5]}$ (LHCb) $\rightarrow$ - 2.08\\
		   	$P_3(B^0\to K^{*0}\mu^+\mu^-)^{[1.1, 2.5]}$ (LHCb) $\rightarrow$ 2.20\\
		   	$P_8^{'\hspace{0.05 cm}}(B^0\to K^{*0}\mu^+\mu^-)^{[1.1, 2.5]}$ (LHCb) $\rightarrow$ - 2.36 \\
		   	$F_H(B^+\to K^+\mu^+\mu^-)^{[2,4.3]}$  (CMS) $\rightarrow$ 2.34\\
			\hline
		\end{tabular}
		\caption{The respective values of pull$^{NP}$ for the additional observables in \textbf{List-2} for the one-operator scenario $\mathcal{O}_9$ with complex WC. The superscripts on the observables indicate the $q^2$ range in {\it GeV}$^2$.}
		\label{tab:pulllst2}
	\end{center}
\end{table}
In table \ref{tab:pulllst2}, the respective pulls for the additional observables in \textbf{List-2} for the one-operator scenario $\mathcal{O}_9$ is shown. For the scenario $\Delta C_9 = - \Delta C_{10}$ also, these observables have similar pulls.

We repeat the fit for the new physics scenario as mentioned above after dropping the 19 data points as listed in \textbf{List-2}\footnote{In this scenario, we perform the analysis with 205 number of data-points which is still larger than that was considered in \cite{Alguero:2019ptt} where they have got a fit-probability of about 50\% for the NP scenario $\Delta C_9=-\Delta C_{10}$.}. The corresponding results are given in the third row of the table \ref{tab:c9eqmc10}. We noticed improvement in the fit quality in terms of the respective $p$-values. 

We have also noted improvement in fit quality for the scenario with a single operator $\mathcal{O}_9$. For example, after dropping the observables listed in `List-2' in the scenario ``Likelihood 2020 dataset", the results are as given below: 
\begin{itemize}
	\item \text{without the CP-asymmetric observables in $B\to K^*$:} $Re(\Delta C_{9})\to -1.08 \pm 0.11,\ Im(\Delta C_{9})\to 1.20 \pm 0.38$,  \text{p-value 95\%}, 
	
	\item \text{with the CP-asymmetric observables in $B\to K^*$:} $Re(\Delta C_{9})\to -1.08 \pm 0.11,\ Im(\Delta C_{9})\to -1.03 \pm 0.48$,  \text{p-value 93.3\%}. 
\end{itemize}
Inspite of the improvement in the respective $p$-values in all the other one operator scenarios, they can not be considered as a good fit. 

\begin{table}[t]
		\renewcommand*{\arraystretch}{1.2}
		\begin{tabular}{|c|c|c|c|}
	     \hline
			Observables from {\bf List-1 or List-2} & Other Experiment & LHCb & NP prediction ($\mathcal{O}_9$)\\
			\hline
			$\text{$BR(B^{0}\to K^{0}\mu^+\mu^-)^{[1, 6]}$}$ (Belle) &  $(0.31\pm 0.19)\times 10^{-7}$ (Belle) & $  (0.92 \pm 0.17) \times 10^{-7}$  &  $(1.12 \pm 0.21) \times 10^{-7}$  \\
			\hline
			$\text{$BR(B^{+}\to K^{+}\mu^+\mu^-)^{[1, 6]}$}$ (Belle) & $(2.3 \pm 0.4) \times10^{-7}$  (Belle)  & $(1.21 \pm 0.07) \times 10^{-7}$  &  $(1.19 \pm 0.22) \times 10^{-7}$  \\
			\hline
			$\text{$A_I(B\to K \mu^+\mu^-)^{[1,6]}$ }$ (Belle)  &  $-0.52\pm 0.19$  (Belle) & N.A.  &  $0.0020\pm 0.0006$  \\
			\hline
			$\text{$S_4^{[4, 6]}$ }$ (ATLAS)  &  $0.32\pm 0.18$ (ATLAS) &  $-0.145 \pm 0.057$  &  $-0.21\pm 0.01$  \\
			\hline
			$\text{$P_4^{'\hspace{0.05 cm}[4, 6]}$ }$ (ATLAS)  &  $0.64\pm 0.38$ (ATLAS) &  $-0.312 \pm 0.116 $  &  $-0.45\pm 0.01$  \\
			
			\hline
			$\text{$P_5^{'\hspace{0.05 cm}[4, 6]}$ }$ (ATLAS) &  $0.26\pm 0.39$ (ATLAS)  &  $-0.439\pm 0.117 $  &  $-0.48\pm 0.05$  \\
			\hline
			$\text{$S_5^{[4, 6]}$ }$ (ATLAS) &  $0.13\pm 0.20$  (ATLAS)   &  $-0.204\pm 0.053$ &  $-0.22\pm 0.03$  \\
			\hline
			$\text{$\frac{dB}{dq^2}(B^{+}\to K^{+}\mu^+\mu^-)^{[0.1, 0.98]}$}$  &  N.A.  & $(0.33 \pm 0.02) \times 10^{-7}$   &  $(0.22 \pm 0.05)\times 10^{-7}$  \\
			\hline
			$\text{$F_H(B^+\to K^+\mu^+\mu^-)^{[2,4.3]}$  }$  &  $0.85\pm 0.35$ (CMS) &  N.A.  &  $0.0225\pm 0.0005$  \\\hline
		\end{tabular}
	\caption{The predicted values of the observables listed in {\bf List-2} (or in {\bf List-1})  for the one operator scenario $\mathcal{O}_9$ which are compared with the correspoding measured values by LHCb and a few other experiments. Similar predictions are obtained in the scenario $\mathcal{O}_9 = - \mathcal{O}_{10}$. }
	\label{tab:predlist1}
\end{table}

In table \ref{tab:predlist1}, we have compared the predicted values of the observables from \textbf{List-2} (or in \textbf{List-1}) in the one operator scenario $\mathcal{O}_9$ with the respective measured values from LHCb and other experiments. This table discusses 9-data points from Lists, out of which eight are measured in experiments other than LHCb. As compared to \textbf{List-1}, in \textbf{List-2} we have added 7 more observables out of which the predicted value of $F_H(B^+\to K^+\mu^+\mu^-)^{[2,4.3]}$ in the NP scenario $\mathcal{O}_9$ has been given in table \ref{tab:predlist1}. Note that the observables $P_6^{'\hspace{0.05 cm}}$, $P_8^{'\hspace{0.05 cm}}$, $P_1$ and $P_3$ are related to $S_7$, $S_8$, $S_3$ and $S_9$, respectively. Therefore, any observed deviations in $P_6^{'\hspace{0.05 cm}}$, $P_8^{'\hspace{0.05 cm}}$, $P_1$ and $P_3$ are related to that in the measured values of $S_7$, $S_8$, $S_3$ and $S_9$, respectively. Based on all these and the discussion above, a few useful remarks about the data points in \textbf{List-2} or \textbf{List-1} are in order:
	\begin{itemize}
		\item We have dropped the two data points $BR(B^{0}\to K^{0}\mu^+\mu^-)^{[1, 6]}$ and $BR(B^{+}\to K^{+}\mu^+\mu^-)^{[1, 6]}$ measured by Belle. As can be seen from table \ref{tab:predlist1}, the measured values of these branching fractions by Belle and LHCb are not consistent with each other, at least within their 1-$\sigma$ error bars. The LHCb data points are included in the analysis. In the new physics scenario, along with the respective parameter spaces favoured by all the analysed data points, the predicted values of the above two branching fractions are in good agreement with that measured by LHCb while these are in tension with Belle. The measured values by Belle have large errors as compared to LHCb. It is essential to mention that many other data points from LHCb on the same branching fractions in different small $q^2$ bins are included in the analysis, which is comfortably explained by the allowed NP scenario. For example, the dataset includes $\frac{dB}{dq^2}(B^{+,0}\to K^{+,0}\mu^+\mu^-)$ from LHCb in all the bins of $q^2$ in between [0.98,6]. Also, we have dropped $\frac{dB}{dq^2}(B^{+}\to K^{+}\mu^+\mu^-)^{[0.1,0.98]}$ by LHCb, as can be seen from the table predicted value is in tension with the measured one. As discussed earlier, this data point has minimal impact on the fit probability (p-value). 
		
		\item The measured value of $A_I(B\to K \mu^+\mu^-)^{[1,6]}$ by Belle has large errors and the allowed NP solution has negligible impact on this observable. 
		
		\item We have presented our main analyses after dropping the data points $P_4'^{[4, 6]}$,  $P_5^{'\hspace{0.05 cm}[4, 6]}$, $S_4^{[4, 6]}$ and $S_5^{[4, 6]}$ measured by ATLAS since these data points have large pulls. The reason is clear from the observation made in table \ref{tab:predlist1}. The data points have large errors and are inconsistent with the relatively precise data from LHCb. The allowed NP solution can easily accommodate these LHCb data points alongside all the other data considered in the analyses. While the respective Belle data largely deviate from the one predicted in the allowed NP scenario.     
		
		\item In the analyses, we have dropped the data points $P_6^{'\hspace{0.05 cm}[1.1, 6]}$, $P_6^{'\hspace{0.05 cm}[4, 6]}$, $S_7^{[1.1, 6]}$ and $S_7^{[4, 6]}$ from LHCb. For the same observables, we include the data points in three other small $q^2$ bins in the region $0.1 \le q^2 \le 4$ $\it GeV^2$. Note that the SM predictions are small, these observables have very low sensitivity to the variation of $\Delta C_9$ \cite{Matias:2012xw,DescotesGenon:2012zf}. We have observed that it is hard to explain the current measured values in the above-mentioned bins for the allowed NP solutions. The data points in the other three bins, as mentioned above, are consistent with the predicted values in the SM. Here, we would like to mention that even though the above four data points impact the statistical significance of the results, if we keep them in the analysis, we get results with allowed $p$-values.
	
	 \item  CMS has measured $F_H(B^+\to K^+\mu^+\mu^-)^{[2,4.3]}$ with an error $\approx$ 40\%.  The corresponding prediction in the NP scenario $\mathcal{O}_9$ is given in table \ref{tab:predlist1} which is consistent with the respective SM prediciton: $(0.0224\pm 0.0004)$. It is hard to explain the current observation by CMS in the allowed NP scenario. Similar arguments hold for the observables $P_1^{[1.1, 2.5]}$, $P_3^{[1.1, 2.5]}$, $P_8^{'\hspace{0.05 cm}[1.1, 2.5]}$, $S_3^{[1.1, 2.5]}$, $S_8^{[1.1, 2.5]}$, and $S_9^{[1.1, 2.5]}$ as these observables are insensitive to $\Delta C_9$ \cite{Altmannshofer:2008dz,Matias:2012xw,DescotesGenon:2012zf}. In the respective bins, the predictions for the allowed NP scenarios do not change from the respective SM predictions. Here, we would like to mention that the measurements are also available in three more bins for each of the observables, and they are all consistent with the respective SM predictions.  
		
	\item Out of the seven observables as mentioned in the last item, $F_H(B^+\to K^+\mu^+\mu^-)^{[2,4.3]}$, $P_8^{'\hspace{0.05 cm}[1.1, 2.5]}$, and  $S_8^{[1.1, 2.5]}$ have major impact in determining the quality of the fit. For example, in the analysis of NP scenario $\mathcal{O}_9$, alongside the data points in \textbf{List-1} if we drop only these three data points (with large pulls) then the $p$-value of the fit increases from 62.5\% (table \ref{tab:oneoprcwclis2}) to 87\%.
		
	\item We have checked that if we drop the Belle and ATLAS data, and $F_H(B^+\to K^+\mu^+\mu^-)^{[2,4.3]}$ from CMS from {\bf List-1}, we get acceptable fit quality $\sim$ 10 $\%$ for the complex $\mathcal{O}_9$ scenario. Therefore, to get an acceptable fit we don't have to drop any LHCb data.
	\end{itemize}

All the observations mentioned in the above items are also applicable for the scenario $\mathcal{O}_9=-\mathcal{O}_{10}$. We can not explain a few data listed above in the allowed NP scenario because of their low sensitivity to this scenario. Also, we notice that to get an acceptable fit, we don't need to drop any LHCb data. Still, the above exercise is helpful for one to see the impact of very few data points in determining the statistical significance of the analysis without changing the outcome. However, we should stress that the data suffer from large error and further conclusions must wait for more precise data.

As mentioned earlier, we present our results based on the most recent `likelihood' data (by LHCb). To understand the trend of the data, we carry out a fit using the old-data (Run-1: Likelihood and moments) on angular observables from LHCb \cite{Aaij:2015oid} with complex WCs. The `Likelihood datasets 2016' and `Moment datasets 2016' contain a total of 259 and 281 data points respectively. For the `moment datasets', including all these observables, one can fit the one-operator scenario $\mathcal{O}_9$ with an allowed $p$-value. However, it is hard to fit the other one-operator scenarios with allowed $p$-values, including the $\Delta C_9 = -\Delta C_{10}$ scenario. On the other hand, for `Likelihood datasets 2016', fit to any NP scenario, including all the observables has a fit-probability $p <<$ 3 $\%$. As discussed earlier, the data points: $P_5^{'}(B^0\to K^{*0}\mu^+\mu^-)^{[4, 6]}$ (ATLAS), $S_5(B^0\to K^{*0}\mu^+\mu^-)^{[4, 6]}$ (ATLAS), $\frac{dB}{dq^2}(B^{+}\to K^{+}\mu^+\mu^-)^{[0.1, 0.98]}$ (LHCb) and $BR(B^{+}\to K^{+}\mu^+\mu^-)^{[1,6]}$ (Belle), and the ones given in table \ref{tab:pullcommon} are common to all the datasets. To be consistent with the earlier analyses, we prepare a third list consisting of these observables only:        
\begin{itemize}	
	\item \textbf{List-3}: It includes $P_5^{'}(B^0\to K^{*0}\mu^+\mu^-)^{[4, 6]}$ (ATLAS), $S_5(B^0\to K^{*0}\mu^+\mu^-)^{[4, 6]}$ (ATLAS), $\frac{dB}{dq^2}(B^{+}\to K^{+}\mu^+\mu^-)^{[0.1, 0.98]}$ (LHCb) and $BR(B^{+}\to K^{+}\mu^+\mu^-)^{[1,6]}$ (Belle) in addition to the observables listed in table \ref{tab:pullcommon} with a pull $>$ 2.5. 
\end{itemize}
After dropping the data points given in \textbf{List-3} from the old-datasets, the fit results for the scenario $\Delta C_9 = -\Delta C_{10}$ is given in the last two rows of  table \ref{tab:c9eqmc10}. Note that in both the datasets, we have results with appreciable fit-probabilities. 

For the one-operator scenarios, the comparative results are shown in Table \ref{tab:comparative} in the appendix. We have dropped the inputs on the observables given in \textbf{List-1} for the ``Likelihood 2020 datasets". Note that in the analyses involving the ``Likelihood 2016 datasets" and ``Moments 2016 datasets" we have dropped the inputs given in \textbf{List-3}. We find that $\mathcal{O}_9$ is the only one operator scenario that can comfortably explain all these data-sets, and a large value of $Im(\Delta C_9)$ is allowed by the current data. The fit with the ``moment-dataset" results in appreciable fit probabilities for all the one operator scenarios, as seen from table \ref{tab:comparative}. The old-data-set includes the CP-asymmetric angular observables in $B\to K^*$ decays while the 2020 update~\cite{Aaij:2020nrf} does not. With a $p$-value $\approx$ 71.6\%, the fit with $\mathcal{O}_9$ for the old Likelihood data is slightly better than the one we obtain for the new-data-set, though the value of the corresponding $\chi^2_{min}$ is larger in the old-dataset as compared to the new dataset. This could be due to the relatively large number of observables in the old datasets. As can be seen from table~\ref{tab:comparative}, the magnitude of the best fit value of $Re(\Delta C_9)$, corresponding to the new dataset reduces by $\approx 10\%$ as compared to the one obtained from the old likelihood-data. 

\begin{table*}[t]
	\scriptsize
	\begin{ruledtabular}
		\renewcommand*{\arraystretch}{1.2}
		\begin{tabular}{ccccccc}
			$\text{Model}$  &  $\text{$\Delta $AICc}$  &    $\text{MSE}_{X-\text{Val}}$  &  $\chi _{\text{Min}}^2\text{/DOF}$  &  $\text{p-val ($\%$)}$  & $\text{Pull}_{\text{SM}}$  & $\text{Result}$  \\
			\hline
			$585$  &  $0.$  &  $0.989$  &  $\text{189.05/206}$  &  $79.6$  &  $9.0$  &  $\begin{array}{l}
			\left.\text{Re(}\text{$\Delta $C}_9\right)\to \text{-1.36$\pm $0.24},~
			\left.\text{Im(}\text{$\Delta $C}_9\right)\to \text{2.05$\pm $0.36} \\
			\left.\text{Re(}C_9'\right)\to \text{0.57$\pm $0.23},~
			\left.\text{Im(}C_9'\right)\to \text{0.14$\pm $0.25} \\
			\left.\text{Re(}\text{$\Delta $C}_{10}\right)\to \text{0.51$\pm $0.22},~
			\left.\text{Im(}\text{$\Delta $C}_{10}\right)\to \text{-0.53$\pm $0.46} \\
			\end{array}$  \\
			\hline
			$18$  &  $2.015$  &  $0.942$  &  $\text{199.41/210}$  &  $68.9$  &  $9.2$  &  $\begin{array}{l}
			\left.\text{Re(}\text{$\Delta $C}_9\right)\to \text{-1.08$\pm $0.099},~
			\left.\text{Re(}C_9'\right)\to \text{0.50$\pm $0.18} \\
			\end{array}$  \\
			\hline
			$697$  &  $3.506$  &  $0.971$  &  $\text{188.25/204}$  &  $77.9$  &  $8.7$  &  $\begin{array}{l}
			\left.\text{Re(}\text{$\Delta $C}_9\right)\to \text{-1.4$\pm $0.24},~
			\left.\text{Im(}\text{$\Delta $C}_9\right)\to \text{1.93$\pm $0.49} \\
			\left.\text{Re(}C_9'\right)\to \text{0.56$\pm $0.23},~
			\left.\text{Im(}C_9'\right)\to \text{0.31$\pm $0.5} \\
			\left.\text{Re(}\text{$\Delta $C}_{10}\right)\to \text{0.52$\pm $0.22},~
			\left.\text{Im(}\text{$\Delta $C}_{10}\right)\to \text{-0.51$\pm $0.41} \\
			\left.\text{Re(}C_{10}'\right)\to \text{-0.032$\pm $0.177},~
			\left.\text{Im(}C_{10}'\right)\to \text{0.75$\pm $0.82} \\
			\end{array}$  \\
			\hline
			$529$  &  $3.791$  &  $0.977$  &  $\text{197.05/208}$  &  $69.6$  &  $8.9$  &  $\begin{array}{l}
			\left.\text{Re(}\text{$\Delta $C}_9\right)\to \text{-1.11$\pm $0.11},~
			\left.\text{Im(}\text{$\Delta $C}_9\right)\to \text{-0.12$\pm $0.46} \\
			\left.\text{Re(}C_9'\right)\to \text{0.42$\pm $0.23},~
			\left.\text{Im(}C_9'\right)\to \text{-1.21$\pm $0.41} \\
			\end{array}$  \\
			\hline	
			$641$  &  $4.946$  &  $1.011$  &  $\text{189.69/204}$  &  $75.6$  &  $8.6$  &  $\begin{array}{l}
			\left.\text{Re(}C_7'\right)\to \text{-0.0075$\pm $0.0136},~
			\left.\text{Im(}C_7'\right)\to \text{-0.015$\pm $0.037} \\
			\left.\text{Re(}\text{$\Delta $C}_9\right)\to \text{-1.07$\pm $0.13},~
			\left.\text{Im(}\text{$\Delta $C}_9\right)\to \text{-0.061$\pm $0.296} \\
			\left.\text{Re(}C_9'\right)\to \text{0.61$\pm $0.25},~
			\left.\text{Im(}C_9'\right)\to \text{-1.98$\pm $0.4} \\
			\left.\text{Re(}\text{$\Delta $C}_{10}\right)\to \text{0.59$\pm $0.21},~
			\left.\text{Im(}\text{$\Delta $C}_{10}\right)\to \text{0.051$\pm $1.254} \\
			\end{array}$  \\
			\hline
			$530$  &  $5.479$  &  $0.993$  &  $\text{198.74/208}$  &  $66.6$  &  $8.8$  &  $\begin{array}{l}
			\left.\text{Re(}\text{$\Delta $C}_9\right)\to \text{-1.34$\pm $0.26},~
			\left.\text{Im(}\text{$\Delta $C}_9\right)\to \text{1.95$\pm $0.44} \\
			\left.\text{Re(}\text{$\Delta $C}_{10}\right)\to \text{0.32$\pm $0.23},~
			\left.\text{Im(}\text{$\Delta $C}_{10}\right)\to \text{-0.56$\pm $0.57} \\
			\end{array}$  \\
			\hline
			$513$  &  $5.492$  &  $0.98$  &  $\text{202.89/210}$  &  $62.5$  &  $9.0$  &  $\begin{array}{l}
			\left.\text{Re(}\text{$\Delta $C}_9\right)\to \text{-1.1$\pm $0.11},~
			\left.\text{Im(}\text{$\Delta $C}_9\right)\to \text{1.27$\pm $0.37} \\
			\end{array}$  \\
		\end{tabular}
	\end{ruledtabular}
	\caption{\small The selected models pass the criterion of $\Delta\text{AIC}_c \leq 6$ and MSE$_{\text{X-val}} < 1.5$ for the Likelihood 2020 dataset after dropping the observables mentioned in List-1. For all the selected models, we have calculated $\text{Pull}_{\text{SM}}$, with $\chi^2_{SM} = 288.9$ and p-value of SM = 0.035 $\%$. The parameter uncertainties are obtained from hessian matrix. The operator(s) involved in a particular scenario (specified by a number) has been described in table \ref{tab:modelopr} in the appendix.}
	\label{tab:ModselNew}
\end{table*}

\begin{table*}[t]
	\scriptsize
	\begin{ruledtabular}
		\renewcommand*{\arraystretch}{1.3}
	\begin{tabular}{ccccccc}
		$\text{Model Index}$  &  $\text{$\Delta $AIC}_c$  &   $\text{MSE}_{X-\text{val}}$&  $\chi _{\text{Min}}^2/\text{DOF}$ & p-val (\%) & $\text{Pull}_{\text{SM}}$  &  $\text{Values}$ \\
		\hline
		$18$  &  $0.$  &  $0.948$  &  $\text{232.7/245}$  &  $70.4$  &  $9.2$  &  $\begin{array}{l}
		\left.\text{Re(}\text{$\Delta $C}_9\right)\to \text{-1.079$\pm $0.099},~
		\left.\text{Re(}C_9'\right)\to \text{0.50$\pm $0.18} \\
		\end{array}$  \\
		\hline
		$585$  &  $1.466$  &  $0.981$  &  $\text{225.87/241}$  &  $75.0$  &  $8.8$  &  $\begin{array}{l}
		\left.\text{Re(}\text{$\Delta $C}_9\right)\to \text{-1.13$\pm $0.11},~
		\left.\text{Im(}\text{$\Delta $C}_9\right)\to \text{-0.0073$\pm $0.2829}\\
		\left.\text{Re(}C_9'\right)\to \text{0.61$\pm $0.23},~
		\left.\text{Im(}C_9'\right)\to \text{1.77$\pm $0.39} \\
		\left.\text{Re(}\text{$\Delta $C}_{10}\right)\to \text{0.46$\pm $0.21},~
		\left.\text{Im(}\text{$\Delta $C}_{10}\right)\to \text{0.23$\pm $0.32} \\
		\end{array}$  \\
		\hline
		$529$  &  $2.310$  &  $0.977$  &  $\text{230.9/243}$  &  $70.1$  &  $8.9$  &  $\begin{array}{l}
		\left.\text{Re(}\text{$\Delta $C}_9\right)\to \text{-1.12$\pm $0.11},~
		\left.\text{Im(}\text{$\Delta $C}_9\right)\to \text{0.0094$\pm $0.4714} \\
		\left.\text{Re(}C_9'\right)\to \text{0.46$\pm $0.22},~
		\left.\text{Im(}C_9'\right)\to \text{1.03$\pm $0.45} \\
		\end{array}$  \\
		\hline
		$697$  &  $4.598$  &  $0.974$  &  $\text{224.74/239}$  &  $73.7$  &  $8.5$  &  $\begin{array}{l}
		\left.\text{Re(}\text{$\Delta $C}_9\right)\to \text{-1.25$\pm $0.16},~
		\left.\text{Im(}\text{$\Delta $C}_9\right)\to \text{-0.098$\pm $0.381} \\
		\left.\text{Re(}C_9'\right)\to \text{0.59$\pm $0.24},~
		\left.\text{Im(}C_9'\right)\to \text{1.78$\pm $0.42} \\
		\left.\text{Re(}\text{$\Delta $C}_{10}\right)\to \text{0.44$\pm $0.22},~
		\left.\text{Im(}\text{$\Delta $C}_{10}\right)\to \text{0.52$\pm $0.43} \\
		\left.\text{Re(}C_{10}'\right)\to \text{-0.063$\pm $0.173},~
		\left.\text{Im(}C_{10}'\right)\to \text{0.26$\pm $0.28} \\
		\end{array}$  \\
		\hline
		$2$  &  $5.032$  &  $0.964$  &  $\text{239.77/246}$  &  $60.0$  &  $9.1$  &  $\begin{array}{l}
		\left.\text{Re(}\text{$\Delta $C}_9\right)\to \text{-1.06$\pm $0.11} \\
		\end{array}$  \\
		\hline
		$513$  &  $5.402$  &  $0.982$  &  $\text{238.1/245}$  &  $61.2$  &  $8.9$  &  $\begin{array}{l}
		\left.\text{Re(}\text{$\Delta $C}_9\right)\to \text{-1.09$\pm $0.11},~
		\left.\text{Im(}\text{$\Delta $C}_9\right)\to \text{-1.11$\pm $0.46} \\
		\end{array}$  \\
		\hline
		$641$  &  $5.430$  &  $0.993$  &  $\text{225.58/239}$  &  $72.4$  &  $8.5$  &  $\begin{array}{l}
		\left.\text{Re(}C_7'\right)\to \text{-0.0059$\pm $0.0134},~
		\left.\text{Im(}C_7'\right)\to \text{-0.0094$\pm $0.0282} \\
		\left.\text{Re(}\text{$\Delta $C}_9\right)\to \text{-1.12$\pm $0.12},~
		\left.\text{Im(}\text{$\Delta $C}_9\right)\to \text{-0.054$\pm $0.309} \\
		\left.\text{Re(}C_9'\right)\to \text{0.66$\pm $0.26},~
		\left.\text{Im(}C_9'\right)\to \text{1.8$\pm $0.41} \\
		\left.\text{Re(}\text{$\Delta $C}_{10}\right)\to \text{0.48$\pm $0.21},~
		\left.\text{Im(}\text{$\Delta $C}_{10}\right)\to \text{0.33$\pm $0.43} \\
		\end{array}$  \\
		\hline
		$20$  &  $5.569$  &  $0.967$  &  $\text{238.27/245}$  &  $60.9$  &  $8.9$  &  $\begin{array}{l}
		\left.\text{Re(}\text{$\Delta $C}_9\right)\to \text{-1.11$\pm $0.11},~
		\left.\text{Re(}C_{10}'\right)\to \text{-0.13$\pm $0.11} \\
		\end{array}$  \\
	\end{tabular}
\end{ruledtabular}
	\caption{\small Same as table \ref{tab:ModselNew}, however, now the inputs from $B\to K^*$ asymmetric observables from the 2016 Likelihood dataset are included. Here also, $\text{Pull}_{\text{SM}}$ $>$ 8 $\sigma$ for all the selected models with $\chi^2_{SM} = 322.1$ and p-value of SM = 0.46 $\%$. The corresponding operator(s) in a particular scenario is given in table \ref{tab:modelopr}.}
	\label{tab:withassy}
\end{table*}

\section{Model selection}

As seen above, ${\mathcal O}_9$ is the only one-operator scenario that can comfortably explain the present `likelihood' data. In the future, more precise data might prefer more complex multi-operator scenarios or models. However, with the increasing complexity of a model, its predictive capability deteriorates. Thus, model selection needs to take both goodness of the fit and the complexity of the competing models into account. Selecting the best model from a set of candidates for a given set of data is thus not an easy task. Several criteria are available in the literature that achieve this. Keeping at par with our earlier publication \cite{Bhattacharya:2019dot}, in order to measure model performance and select the best model from a set of potential models we use the mean squared error (MSE) and small-sample-corrected Akaike’s Information Criterion (AIC$c$), which is defined as 
\begin{equation}\small
{\rm AIC}c = \chi^2_{min} + 2 K + \frac{2 K (K+1)}{n - K -1}\,,
\label{eq:aicc}
\end{equation}
where $n$ is the sample-size and $K$ is the number of estimable parameters. The applicability of these criteria and other details are provided in~\cite{Bhattacharya:2019dot}.  

For model selection, `cross-validation' (CV) is the most generally applicable, powerful, reliable, and computationally expensive method. The most straightforward and  expensive version of cross-validation is ``leave-one-out cross-validation" (LOOCV). In LOOCV, one of the data points is left out and the rest of the sample (``training set'') is optimized for a particular model. Then that result is used to find the predicted squared error (SE) for the left out data point, which is defined for the $i^{th}$ observable as given below
\begin{equation}
SE = \frac{(\mathcal{O}^{exp}_i-\mathcal{O}^{theory}_i)^2}{\sqrt{(\sigma_{exp}^2+\sigma_{theory}^2)}},
\end{equation}
where $\sigma_{exp}$ and $\sigma_{theory}$ are the experimental and theory errors of the the $i^{th}$ observable. This process is repeated for all data points and a mean-squared-error (MSE) is obtained using all those residuals. This process is repeated for all models. The models with the least MSE are the best ones. Although leave-one-out-cross-validation (LOO-CV) is asymptotically equivalent to AICc, there are differences between the two. Theoretical considerations aside, AICc is just likelihood penalized by the degrees of freedom. Evidently, AICc accounts for uncertainty in the data (-2Log(L)) and makes the assumption that more parameters leads to higher risk of over-fitting (2k). Cross-validation (CV) just looks at the test set performance of the model, with no further assumptions. There is no explicit measure of model complexity, unlike AICc. Clearly, AICc penalizes model complexity more than CV. The accepted practice in the literature is that if one cares mostly about making predictions and assumes the test set(s) to be reasonably similar to the validation sets, one should go for CV (only with large number of data).

Following the reasoning in the paragraph above, it seems clear \textit{a priori} that models with low MSE but not selected by AICc may have more parameters. Furthermore, this study clearly finds that under the simultaneous application of both AICc and CV, the models cluster in such a way that results in a more robust selection of models than the use of any one of these criteria alone. So the fact that some of the selected models by CV are further discarded by AIC due to relative complexity is actually quite a non-trivial finding.

\begin{figure*}[t]
	\centering
	\subfloat[]
	{\includegraphics[width=0.45\textwidth]{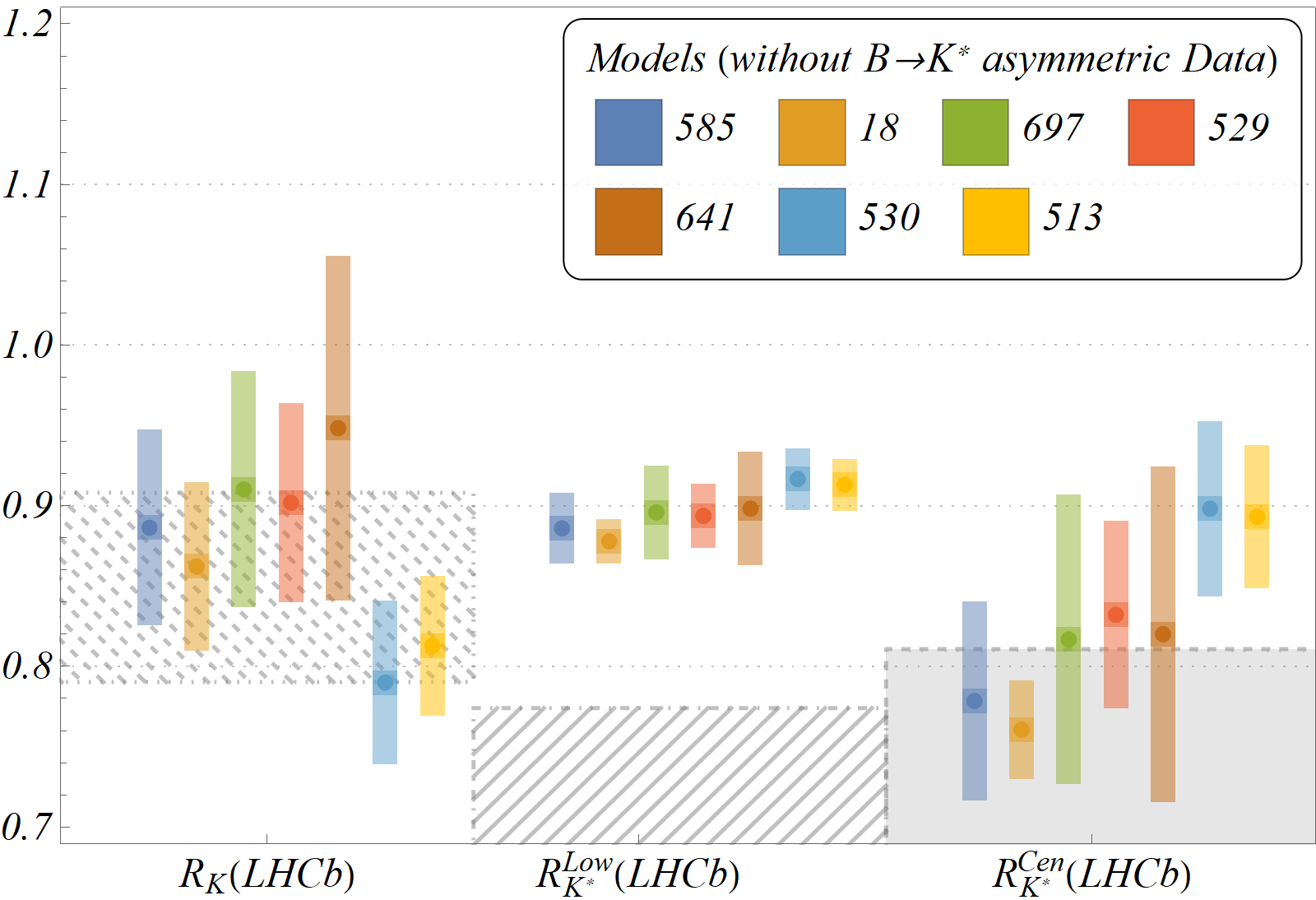}\label{fig:RKRKstq2}~~~~
\includegraphics[width=0.45\textwidth]{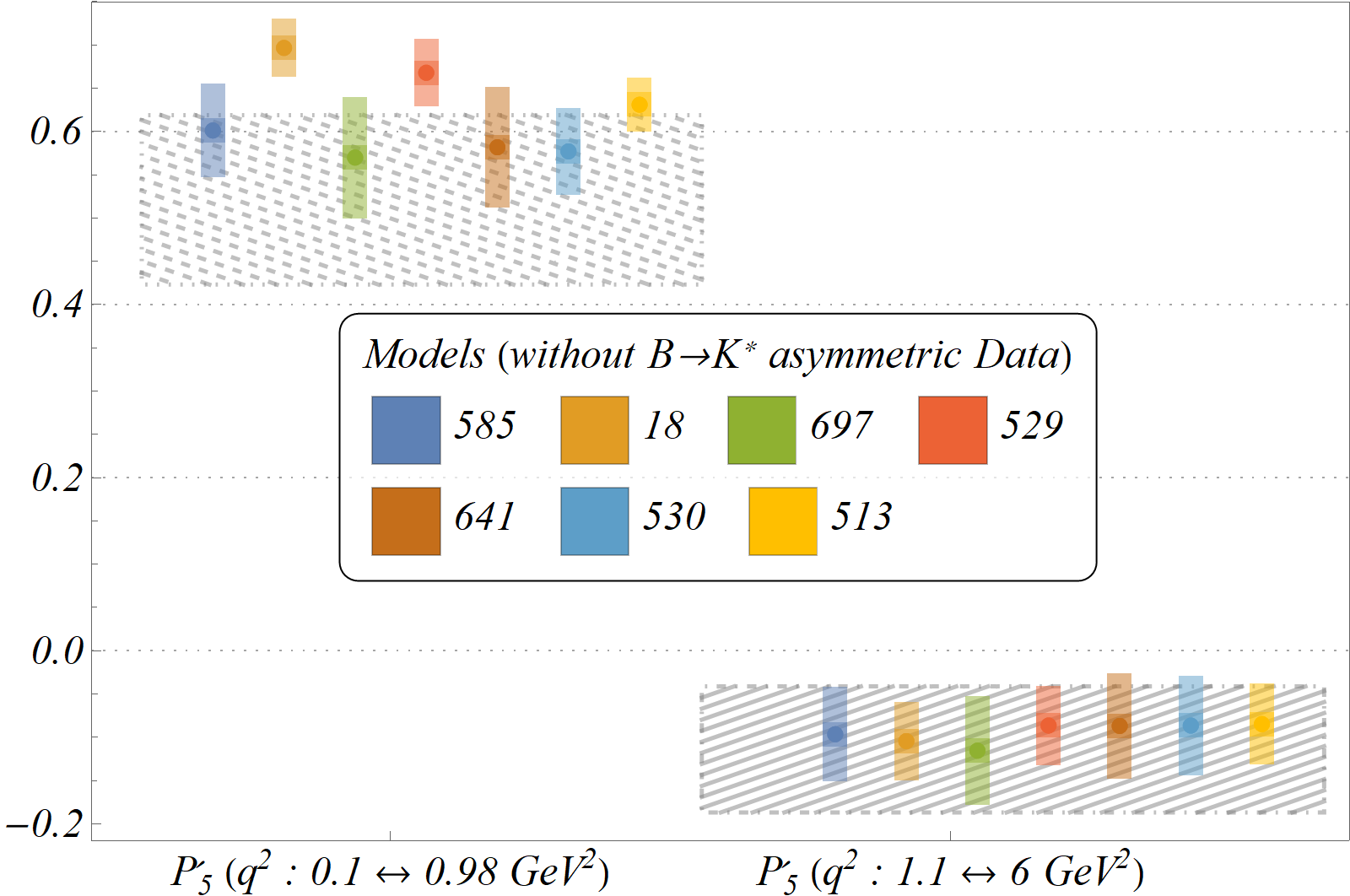}\label{fig:P5prq2}}\\
	\subfloat[]
	{\includegraphics[width=0.45\textwidth]{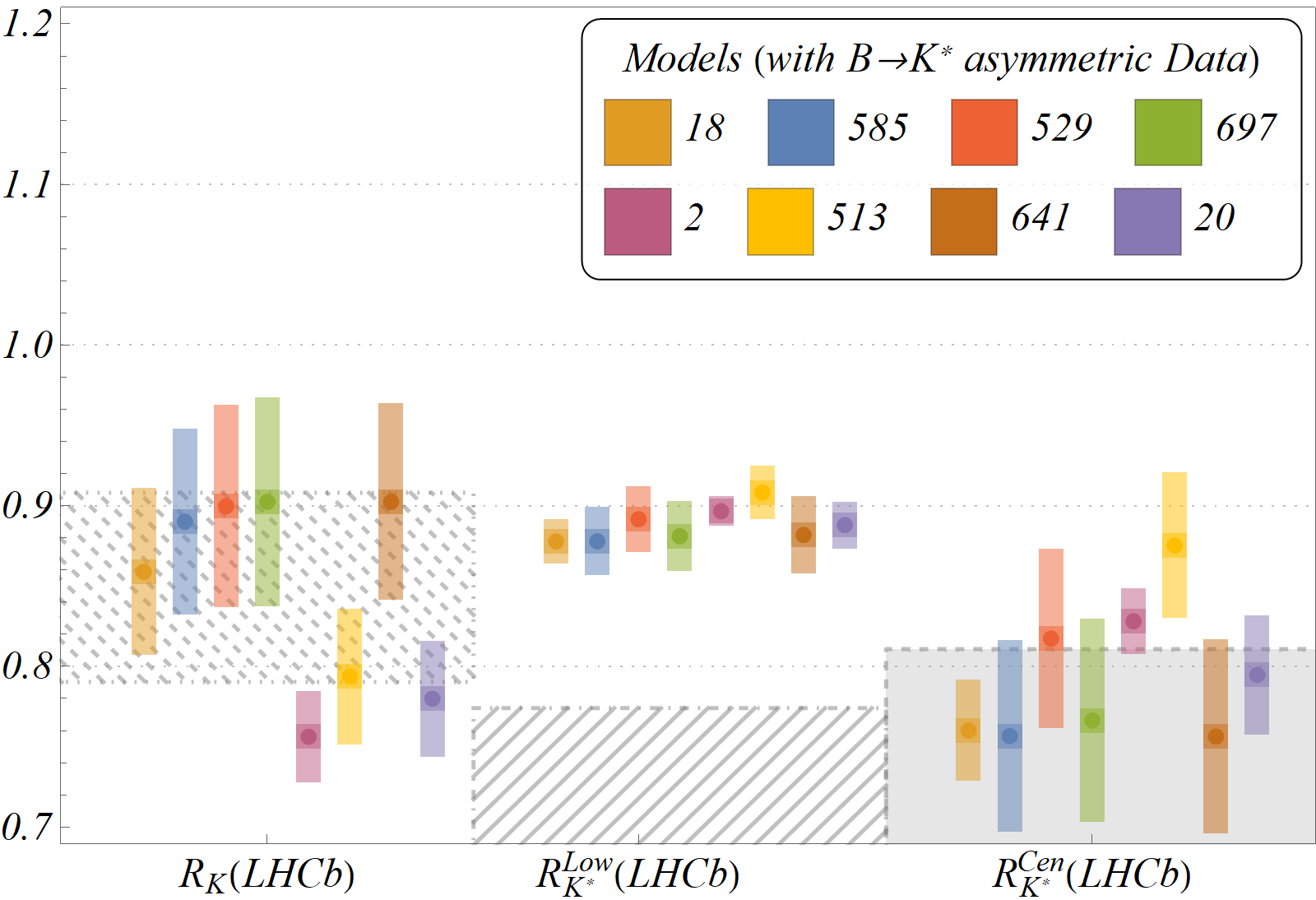}\label{fig:RKRKstq2wassym}~~~~
	\includegraphics[width=0.45\textwidth]{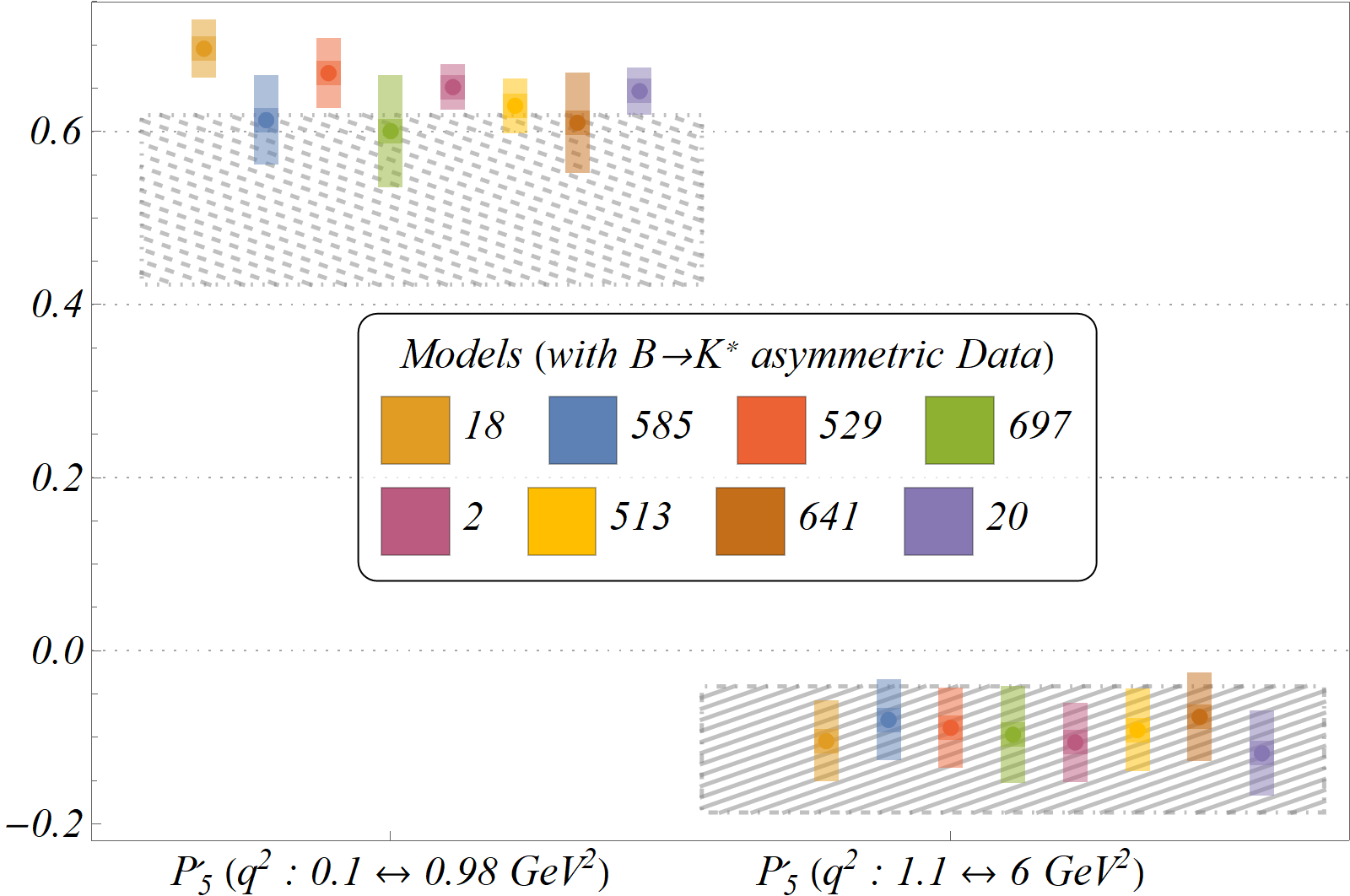}\label{fig:P5prq2w1ssym}}\\
	\caption{\small (a) Predictions of $R_K^{(*)}$ and $P'_5$ in the different $q^2$ regions in our selected models given in table \ref{tab:ModselNew}. Shaded regions are corresponding experimental $1\sigma~ CL$s. (b) Same as fig. \ref{fig:RKRKstq2}  for the selected models given in table \ref{tab:withassy}. }
	\label{fig:Rkpredictions}
\end{figure*}

As shown from eq. \ref{eq:effopr}, there are 9 operators and corresponding WCs that we are considering. Considering all of them to be real, there are $\sum_{k=1,9} (9!/k!(9-k)!) = 511$ possible combinations, i.e. scenarios to be considered. If we consider the WCs to be complex in general, disregarding the combinations with a mixture of real and complex WCs, there are exactly those many cases where the WCs are complex. In total, there are 1022 possible combinations of the coefficients forming a predefined global set of different scenarios. Thus, the considered scenarios only include cases where the WCs are either all real or all complex. Among several competing models, we select the best ones that explain the data by using the conservative limit of $\Delta\text{AIC}c \leq 6$ and with a low value of MSE$_{\text{X-val}} < 1.5$. There is no particular reason for choosing the allowed values of MSE$_{\text{X-val}}$ lower than 1.5, it is just a choice. The idea is to check that the selected scenarios in AIC$_c$ should not have a very high MSE score. Also, we have noted that most of the models which are not selected by both the criteria have MSE score $>> 1.5$. Out of these 1022 possible combinations, only a few are selected by these criteria and are listed in table \ref{tab:ModselNew}. We rank the selected models according to the value of $\Delta\text{AIC}c$. Note that $\mathcal{O}_9$ with complex WC, though not the best model, is the only one-operator scenario passing all the selection criteria. Some two, three and four-operator scenarios are selected, and all of these contain $\mathcal{O}_9$ (with real or complex WC) as one of the operators. Note that apart from model 18, all the selected models have complex WCs. Here, we have dropped a few selected complex models with five operator scenarios. The best model is a three-operator scenario: [$\mathcal{O}_9$, $\mathcal{O'}_9$, $\mathcal{O}_{10}$] with complex WCs. As expected, the real parts of the WCs are large. Also a large non-zero contribution from $Im(\Delta C_9)$ is required in all the selected cases. In a few cases large non-zero contributions from $Im( C'_{9})$ is also allowed. Note that we have obtained the above results in the absence of the CP-asymmetric observables from $B\to K^*$ decays. We have explicitely checked that this conclusion holds even if we drop the asymmetric observables in $B_s\to\phi\mu\mu$ decays. For this particular dataset $\chi^2_{SM} = 288.9$.

As we have discussed earlier, our main datasets due to `Likelihood 2020' do not contain the CP-asymmetric observables of $B \to K^*$ decay modes. It is expected that these observables will be more sensitive to the imaginary parts of the WCs. As shown in table \ref{tab:oneoprcwclis2}, we have carried out a fit after adding the data on these observables from the `Likelihood 2016 datasets' with those in `Likelihood 2020 datasets'. The results show that the allowed range of $Re(\Delta C_9)$ remains the same, while the $Im(\Delta C_9)$ changes its sign because of a few asymmetric observables as shown earlier. We have checked that almost all the models given in table \ref{tab:ModselNew} are selected even after incorporating the additional datasets as mentioned above. The corresponding details are provided in table \ref{tab:withassy}. There are a few changes in the ranking of the respective models. Now in the AIC$_c$ ranking, the two operator scenario $[\mathcal{O}_9, \mathcal{O}^{'}_{9}]$ with real WCs is slightly ahead of the three operator scenario $[\mathcal{O}_9, \mathcal{O}_{10}, \mathcal{O}^{'}_{9}]$ (model-585) with complex WCs. This is due to a slight increase in the value of $\chi^2_{min}/DOF$ for the model-585. There are a few more complex models which are selected for $4 < \text{$\Delta $AIC}_c < 6$ but are not listed in this table. The one operator scenario $\mathcal{O}_9$ with real WC appears in the list with a value of $\text{$\Delta $AIC}_c = 5.032$ and $\text{MSE}_{X-\text{val}} = 0.964$, respectively. The same scenario has a relatively high value of $\text{$\Delta $AIC}_c = 8.51$ in the analysis without the CP-asymmetric data in $B\to K^*$ modes. Also, the two operator scenario $[\mathcal{O}_9, \mathcal{O}^{'}_{10}]$ with real WCs is now selected by the data with $\text{$\Delta $AIC}_c = 5.569$ and $\text{MSE}_{X-\text{val}} = 0.967$. Earlier this model was not selected because of its high value of $\text{$\Delta $AIC}_c$. On the other hand, the two operator scenario $[\mathcal{O}_9, \mathcal{O}_{10}]$ (model no. $530$) with the complex WCs is not selected by the present datasets since it has $\text{$\Delta $AIC}_c = 7.32$, though the associated $p$ value is 61.58\%. 

We note from table \ref{tab:withassy} that in all the two or more operator scenarios, the two operators $\mathcal{O}_9$ and $\mathcal{O}^{'}_9$ are common and $Im(C^{'}_9)$ has a sizable positive contribution that is not consistent with zero at 68\% confidence interval (CI). However, $Im(\Delta C_9)$ is consistent with zero even at 68\% CI, though it could be large within the allowed interval since the estimated error is large. The differences in the allowed CIs of $Im(\Delta C_9)$ between the results presented in tables \ref{tab:ModselNew} and \ref{tab:withassy} could be because a large positive value of $Im(\Delta C_9)$ is not allowed by the CP-asymmetric data in $B\to K^*$ modes (see Figs. \ref{fig:A801098}, \ref{fig:A81125}, \ref{fig:A8116} and \ref{fig:A8254}). Similarly, in the relevant scenarios $Im(\Delta C_{10})$ or $Im(C^{'}_{10})$ could be large while being consistent with zero. 

Fig. \ref{fig:RKRKstq2} shows the predictions of $R_K^{(*)}$ and $P'_5$ in different $q^2$ bins for our selected models from table \ref{tab:ModselNew}, while comparing them with the corresponding measurements. Note that here the correlations between the SM and the NP parameters have been neglected. In all the models, the predictions are consistent with the measurements of $R_K^{(*)}$ by Belle. Our best model can accommodate all the observed data on $R_K^{(*)}$ and $P'_5$ at 1-$\sigma$ CL, except the $R^{low}_{K^*}(LHCb)$.  In fact, none of the selected models can explain the $R^{low}_{K^*}(LHCb)$ at 1-$\sigma$. At low-$q^2$ the $B\to K^*\ell\ell$ decay rates are dominated by $1/q^2$ enhanced photon contributions and as pointed out earlier in several model independent analyses \cite{Altmannshofer:2017yso,Capdevila:2017bsm,Geng:2017svp}, a NP contact interaction that explains $R_K$ and $R_{K^*}^{Cen}$ affects $R_{K^*}^{Low}$  typically by at most 10\%. Also, the possibility of explaining $R_{K^*}^{Low}$ with tensor operators has been addressed in ref. \cite{Bardhan:2017xcc}, which we are not considering in this analysis (see also \cite{Altmannshofer:2017bsz}). Note that in all the selected models, the predicted results are consistent with their respective measured values at 2-$\sigma$. In model 641, the predictions of $R_{K}$ and $R_{K^*}^{Cen}$ have relatively large errors compared to that in $R_{K^*}^{Low}$. This is due to the large errors in the fit result of $Im(\Delta C_{10})$, while $R_{K^*}^{Low}$ is very less sensitive to these WCs. Similar predictions/comparisons are provided in Fig. \ref{fig:RKRKstq2wassym} for the selected models in table \ref{tab:withassy}. No noticeable changes are observed except now the errors in $R_{K}$ and $R_{K^*}^{Cen}$ have reduced because of the reason mentioned above.  

\section{Summary}

Following a model-independent effective theory approach with dimension-six operators, we have analyzed the new physics effects in $b \to s\ell\ell$ decays, based on the data available till date. To the best of our knowledge, we are analyzing the relevant operator basis with complex WCs for the first time in literature. We have found that $\mathcal{O}_9$ is the only one-operator scenario with both real and complex WC (with a large non-zero imaginary part), which can provide a plausible explanation of the given data. This is the case even if all the CP-asymmetric observables are dropped from the fit. We have pointed out the corresponding CP-averaged and CP-asymmetric observables which could be the probable source of such large imaginary contributions. Given the data, we have used the method of model selection incorporating both AIC$_c$ and cross-validation to pinpoint the best possible combination of operators with real and complex WC, which can best explain the data. The scenario with $\mathcal{O}_9$ is the only one-operator scenario which passes the test. However, there are a few two, three and four-operator scenarios which have passed all the criteria set by the selection methods. Allowed confidence intervals of the new WCs are shown. For the selected models, we have provided predictions for various observables and compared them. 
\vskip 0.5cm

{\bf Acknowledgments:} This work of S.N. is supported by the Science and Engineering Research Board, Govt. of India, under the grant CRG/2018/001260.


\setcounter{secnumdepth}{0}
\section{Appendix}
In our main results, we have shown the analysis with the most recent data which includes the updated measurements of angular observables corresponding to an integrated luminosity of 4.7 $fb^{-1}$ of $pp$ collision data collected with the LHCb experiment during the years 2011, 2012 and 2016 \cite{Aaij:2020nrf}. This new analysis of LHCb includes the data from Run-1 (2011, 2012) \cite{Aaij:2015oid}, hence, we have not added it separately in our analysis along with this new data. However, we have compared these results with those obtained from an analysis of the LHCb Run-1 data \cite{Aaij:2015oid}. This new data is based on `likelihood' analysis. However, the Run-1 results contain two different data sets which are based on `likelihood' and `moment' data analysis, respectively. For a comparative study, we have carried out the analysis over all these sets. The results are presented in table \ref{tab:comparative}. 

\begin{table*}[htbp]
	\scriptsize
	\begin{ruledtabular}
		\renewcommand*{\arraystretch}{1.2}
		\begin{tabular}{c|ccc|ccc}
			$\text{Dataset}$  &  $\chi ^2\text{/DOF}$  &  $\text{p-val($\%$)}$  &  $\text{Value}$  &  $\chi ^2\text{/DOF}$  &  $\text{p-val($\%$)}$  &  $\text{Value}$  \\
			\cline{2-7}
			&  \multicolumn{3}{c|}{$C_7^{'}$}  &  \multicolumn{3}{c}{$\Delta C_9$}  \\
			\hline
			Likelihood 2020  &  $\text{279.5/210}$  &  $9.4\times 10^{-2}$  &  $\begin{array}{l}
				Re(C_7^{'} )\to \text{-0.039$\pm$0.013} \\
				Im(C_7^{'} )\to \text{-0.026$\pm$0.101} \\
			\end{array}$  &  $\text{202.89/210}$  &  $62.5$  &  $\begin{array}{l}
				Re(\Delta C_{9})\to \text{-1.10$\pm$0.11} \\
				Im(\Delta C_{9})\to \text{1.27$\pm$0.37} \\
			\end{array}$  \\
			\hline
			Likelihood 2016  &  $\text{306.45/249}$  &  0.7  &  $\begin{array}{l}
				Re(C_7^{'} )\to \text{ -0.03$\pm$0.01} \\
				Im(C_7^{'} )\to \text{ -0.002$\pm$0.025} \\
			\end{array}$  &  $\text{235.79/249}$  &  $71.6$  &  $\begin{array}{l}
				Re(\Delta C_{9})\to \text{ -1.21$\pm$0.14} \\
				Im(\Delta C_{9})\to \text{ -1.25$\pm$0.44} \\
			\end{array}$  \\
			\hline
			Moments 2016  &  $\text{291.85/271}$  &  $18.4$  &  $\begin{array}{l}
				Re(C_7^{'} )\to \text{-0.031$\pm$0.016} \\
				Im(C_7^{'} )\to \text{-0.0057$\pm$0.0300} \\
			\end{array}$  &  $\text{241.7/271}$  &  $89.9$  &  $\begin{array}{l}
				Re(\Delta C_{9})\to \text{-1.24$\pm$0.18} \\
				Im(\Delta C_{9})\to \text{1.19$\pm$0.48} \\
			\end{array}$  \\
			\hline
			&  \multicolumn{3}{c|}{$C_9^{'}$}  &  \multicolumn{3}{c}{$\Delta C_{10}$}  \\
			\hline
			Likelihood 2020  &  $\text{287.9/210}$  &  $2.9\times 10^{-2}$  &  $\begin{array}{l}
				Re(C_9^{'})\to \text{ -0.077$\pm$0.149} \\
				Im(C_9^{'})\to \text{ -0.70$\pm$0.54} \\
			\end{array}$  &  $\text{276.17/210}$  &  0.15  &  $\begin{array}{l}
				Re(\Delta C_{10})\to \text{0.64$\pm$0.18} \\
				Im(\Delta C_{10})\to \text{1.79$\pm$0.29} \\
			\end{array}$  \\
			\hline
			Likelihood 2016  &  $\text{310.72/249}$  &  0.47  &  $\begin{array}{l}
				Re(C_9^{'})\to \text{-0.13$\pm$0.15} \\
				Im(C_9^{'})\to \text{-0.15$\pm$0.71} \\
			\end{array}$  &  $\text{303.22/249}$  &  1.1  &  $\begin{array}{l}
				Re(\Delta C_{10})\to \text{ 0.39$\pm$0.15} \\
				Im(\Delta C_{10})\to \text{ 0.45$\pm$0.50} \\
			\end{array}$  \\
			\hline
			Moments 2016  &  $\text{295.4/271}$  &  $14.8$  &  $\begin{array}{l}
				Re(C_9^{'})\to \text{-0.060$\pm$0.148} \\
				Im(C_9^{'})\to \text{-0.084$\pm$0.423} \\
			\end{array}$  &  $\text{281.7/271}$  &  $31.5$  &  $\begin{array}{l}
				Re(\Delta C_{10})\to \text{0.51$\pm$0.14} \\
				Im(\Delta C_{10})\to \text{-0.11$\pm$0.68} \\
			\end{array}$  \\
			\hline
			&  \multicolumn{3}{c|}{$C_{10}^{'}$}  &  \multicolumn{3}{c}{$C_S$}  \\
			\hline
			Likelihood 2020  &  $\text{278.1/210}$  & 0.1  &  $\begin{array}{l}
				Re(C_{10}^{'})\to \text{ 0.33$\pm$0.11} \\
				Im(C_{10}^{'})\to \text{ -0.21$\pm$0.81} \\
			\end{array}$  &  $\text{288.55/210}$  &  $2.6\times 10^{-2}$  &  $\begin{array}{l}
				Re(C_S)\to \text{ -0.029$\pm$0.483} \\
				Im(C_S)\to \text{ -0.032$\pm$0.440} \\
			\end{array}$  \\
			\hline
			Likelihood 2016  &  $\text{303.0/249}$  &  1.1  &  $\begin{array}{l}
				Re(C_{10}^{'})\to \text{ 0.33$\pm$0.11} \\
				Im(C_{10}^{'})\to \text{ 0.02$\pm$0.28} \\
			\end{array}$  &  $\text{311.19/249}$  &  0.4  &  $\begin{array}{l}
				Re(C_S)\to \text{ -0.04$\pm$0.04} \\
				Im(C_S)\to \text{ 0.0017$\pm$0.3043} \\
			\end{array}$  \\
			\hline
			Moments 2016  &  $\text{290.2/271}$  &  $20.2$  &  $\begin{array}{l}
				Re(C_{10}^{'})\to \text{0.28$\pm$0.12} \\
				Im(C_{10}^{'})\to \text{-0.0030$\pm$0.3175} \\
			\end{array}$  &  $\text{295.4/271}$  &  $14.8$  &  $\begin{array}{l}
				Re(C_S)\to \text{ -0.027$\pm$0.279} \\
				Im(C_S)\to \text{ 0.030$\pm$0.251} \\
			\end{array}$  \\
			\hline
			&  \multicolumn{3}{c|}{$C_P$}  &  \multicolumn{3}{c}{$C_S^{'}$}  \\
			\hline
			Likelihood 2020  &  $\text{288.52/210}$  &  $2.6\times 10^{-2}$  &  $\begin{array}{l}
				Re(C_P)\to \text{ -0.0075$\pm$0.0135} \\
				Im(C_P)\to \text{ 0.003$\pm$0.241} \\
			\end{array}$  &  $\text{288.51/210}$  &  $2.6\times 10^{-2}$  &  $\begin{array}{l}
				Re(C_S^{'})\to \text{-0.044$\pm$0.053} \\
				Im(C_S^{'})\to \text{0.0055$\pm$0.3001} \\
			\end{array}$  \\
			\hline
			Likelihood 2016  &  $\text{311.22/249}$  & 0.4  &  $\begin{array}{l}
				Re(C_P)\to \text{ -0.0047$\pm$0.1564} \\
				Im(C_P)\to \text{ 0.02$\pm$0.85} \\
			\end{array}$  &  $\text{311.22/249}$  &  0.4  &  $\begin{array}{l}
				Re(C_S^{'})\to \text{ -0.04$\pm$0.17} \\
				Im(C_S^{'})\to \text{ -0.01$\pm$0.62} \\
			\end{array}$  \\
			\hline
			Moments 2016  &  $\text{295.2/271}$  &  $15$  &  $\begin{array}{l}
				Re(C_P)\to \text{0.26$\pm$0.12} \\
				Im(C_P)\to \text{-0.019$\pm$0.847} \\
			\end{array}$  &  $\text{295.4/271}$  &  $14.8$  &  $\begin{array}{l}
				Re(C_S^{'})\to \text{-0.035$\pm$0.157} \\
				Im(C_S^{'})\to \text{-0.020$\pm$0.263} \\
			\end{array}$  \\
			\hline
			&  \multicolumn{3}{c|}{$C_P^{'}$}  &   &   &   \\
			\cline{1-4}
			Likelihood 2020  &  $\text{288.49/210}$  &  $2.6\times 10^{-2}$  &  $\begin{array}{l}
				Re(C_P^{'})\to \text{ 0.0078$\pm$0.0125} \\
				Im(C_P^{'})\to \text{ -0.002$\pm$0.182} \\
			\end{array}$  &   &   &   \\
			\cline{1-4}
			Likelihood 2016  &  $\text{311.2/249}$  &  0.4 &  $\begin{array}{l}
				Re(C_P^{'})\to \text{ 0.007$\pm$0.013} \\
				Im(C_P^{'})\to \text{ -0.0027$\pm$0.2648} \\
			\end{array}$  &   &   &   \\
			\cline{1-4}
			Moments 2016  &  $\text{295.4/271}$  &  $14.8$  &  $\begin{array}{l}
				Re(C_P^{'})\to \text{0.0061$\pm$0.0135} \\
				Im(C_P^{'})\to \text{0.0021$\pm$0.3384} \\
			\end{array}$  &   &   &   \\
		\end{tabular}
	\end{ruledtabular}
	\caption{\small Comparitive study of the results obtained from the analysis of different data set. The results are presented only for the complex Wilson coefficients (WC). For the `Liklihood 2020 datasets', we have dropped the data points given in \textbf{List-1}, while in the analyses involving the `Likelihood 2016 datasets' and `Moments 2016 datasets', we have dropped the inputs given in \textbf{List-3}.}
	\label{tab:comparative}
\end{table*}
The different operator(s) involved in the selected models in tables \ref{tab:ModselNew} and \ref{tab:withassy} are given in table \ref{tab:modelopr}. There are in total 1022 scenarios and it is very difficult to mention all of them which is a tedious job, we are mentioning only those which are selected by our analysis.
\begin{table}
	\begin{tabular}{c|c}
		Model number & Exact Scenario \\
		\hline
		2  & One-operator: $\mathbb{O}_9$ with real WC \\
		\hline
		18 & Two-operator: $[\mathbb{O}_9,\mathbb{O}_9']$ with real WCs  \\
		\hline 
		20 & Two-operator: $[\mathbb{O}_9,\mathbb{O}_{10}']$ with real WCs \\
		\hline 
		513 & One-operator: $\mathbb{O}_9$ with complex WC    \\
		\hline 
		529 & Two-operator: $[\mathbb{O}_9,\mathbb{O}_9']$ with complex WCs  \\
		\hline 
		530 & Two-operator: $[\mathbb{O}_9,\mathbb{O}_{10}]$ with complex WCs \\
		\hline 
		585 & Three-operator: $[\mathbb{O}_9,\mathbb{O}_9',\mathbb{O}_{10}]$ with complex WCs \\
		\hline 
		641 & Four-operator: $[\mathcal{O}_7',  \mathbb{O}_9,\mathbb{O}_9',\mathbb{O}_{10}]$ with complex WCs   \\
		\hline 
		697 & Four-operator: $[\mathbb{O}_9,\mathbb{O}_9',\mathbb{O}_{10}, \mathcal{O}_{10}']$ with complex WCs    \\	
		\hline	
	\end{tabular}
	\caption{The relevant operators in a selected model specified by a number. }
	\label{tab:modelopr}
\end{table}

\newpage

\bibliography{ASIS}

\end{document}